{}%disable hyper at arxiv
{}%use pdflatex at arxiv

\documentclass[11pt,a4paper]{article}

\newif\ifpublic\publictrue
\newif\ifworking\workingfalse
\usepackage{amsfonts,amsbsy,euscript,mathrsfs,fixmath}
\usepackage{amsmath,amssymb,mathtools,arydshln,faktor,slashed}
\usepackage{scalerel}% for tiny indices
\usepackage{dsfont}
\let\mathbb\mathds
\usepackage[nosort]{cite}
\usepackage{graphicx}
\usepackage[usenames,dvipsnames]{color}
\usepackage[normalem]{ulem}
\usepackage{pdfpages}
\definecolor{linkcolor}{rgb}{0,0,0.6}
\usepackage[colorlinks=true,pdfstartview=FitV,linkcolor=linkcolor,
citecolor=linkcolor,urlcolor=linkcolor,hyperindex=true,hyperfigures=false]{hyperref}
\ifpublic\else\usepackage{showkeys}\fi

\def\showkeysrefformat#1{{\normalfont\tiny\ttfamily#1}}
\makeatletter
\def\SK@@ref#1>#2\SK@{%
{\@inlabelfalse\leavevmode\vbox to\z@{%
\vss\SK@refcolor\rlap{\vrule\raise .75em%
\hbox{\showkeysrefformat{#2}}}}}}
\makeatother

\usepackage[a4paper,text={160mm,247mm},centering]{geometry}
\global\arraycolsep=2pt

\allowdisplaybreaks[3]

\numberwithin{equation}{section}

\expandafter\def\expandafter\bfseries\expandafter{\bfseries\ifmmode\else\boldmath\fi}
\expandafter\def\expandafter\mdseries\expandafter{\mdseries\ifmmode\else\unboldmath\fi}
\expandafter\def\expandafter\normalfont\expandafter{\normalfont\ifmmode\else\unboldmath\fi}

\def\beq{\begin{equation}}
\def\eeq{\end{equation}}
\def\beqz{\begin{equation*}}
\def\eeqz{\end{equation*}}
\def\bea{\begin{eqnarray}}
\def\eea{\end{eqnarray}}
\def\bse{\begin{subequations}}
\def\ese{\end{subequations}}

\newcommand{\id}{\mathrm{id}}
\newcommand{\grade}[1]{{(#1)}}
\DeclareMathOperator{\Ad}{Ad}

\def\id{\protect{{1 \kern-.28em {\rm l}}}}

\def\mg{\mathfrak{g}}
\def\STr{\textrm{STr}}
\def\Tr{\textrm{Tr}}

\def\muLR{\mu_{\scriptscriptstyle L,R}}

\def\vk{\varkappa}

\def\mRg{\mathcal{R}_{g}}

\def\L{\scriptscriptstyle L}
\def\R{\scriptscriptstyle R}

\def\EL{\eta_{\scriptscriptstyle L}}
\def\ER{\eta_{\scriptscriptstyle R}}
\def\ELR{{\eta_{\scriptscriptstyle L,R}}}

\def\tE{\tilde{\eta}}
\def\tEL{\tilde{\eta}_{\scriptscriptstyle L}}
\def\tER{\tilde{\eta}_{\scriptscriptstyle R}}
\def\tELR{\tilde{\eta}_{\scriptscriptstyle L,R}}

\def\b{\textrm{b}}

\def\wOmL{\widetilde{O}_{-}^{\scriptscriptstyle L}}
\def\wOmR{\widetilde{O}_{-}^{\scriptscriptstyle R}}

\def\wOmgm{\widetilde{\Omega}_{-}}

\def\wOmgpm{\widetilde{\Omega}_{\pm}}

\def\mOp{\mathcal{O}_{+}}
\def\mOm{\mathcal{O}_{-}}
\def\mOpm{\mathcal{O}_{\pm}}

\def\wmOp{\widetilde{\mathcal{O}}_{+}}
\def\wmOm{\widetilde{\mathcal{O}}_{-}}
\def\wmOpm{\widetilde{\mathcal{O}}_{\pm}}

\def\mSm{\mathcal{S}_{-}}

\def\mE{\mathcal{E}}

\def\mZ{\mathcal{Z}}

\def\mL{\mathcal{L}}

\def\mX{\mathcal{X}}
\def\mY{\mathcal{Y}}
\def\mR{\mathcal{R}}

\def\hTp{{\mathcal{T}}_{+}}
\def\hTm{{\mathcal{T}}_{-}}
\def\hTpm{{\mathcal{T}}_{\pm}}

\def\mJ{\mathcal{J}}
\def\mJp{\mathcal{J}_{+}}
\def\mJm{\mathcal{J}_{-}}
\def\mJpm{\mathcal{J}_{\pm}}

\def\mQp{\mathcal{Q}_{+}}
\def\mQm{\mathcal{Q}_{-}}
\def\mQpm{\mathcal{Q}_{\pm}}

\def\wmQm{\widetilde{\mathcal{Q}}_{-}}
\def\wmQpm{\widetilde{\mathcal{Q}}_{\pm}}

\def\mAp{\mathcal{A}_{+}}
\def\mAm{\mathcal{A}_{-}}
\def\mApm{\mathcal{A}_{\pm}}
\def\mAmp{\mathcal{A}_{\mp}}

\def\wmAp{\widetilde{\mathcal{A}}_{+}}
\def\wmAm{\widetilde{\mathcal{A}}_{-}}
\def\wmApm{\widetilde{\mathcal{A}}_{\pm}}
\def\wmAmp{\widetilde{\mathcal{A}}_{\mp}}

\def\gL{g_{\scriptscriptstyle L}}
\def\gR{g_{\scriptscriptstyle R}}
\def\gLR{g_{\scriptscriptstyle L,R}}

\def\mRg{\mathcal{R}_{g}}

\def\XL{X^{\scriptscriptstyle L}}
\def\XR{X^{\scriptscriptstyle R}}

\def\mKpm{\mathcal{K}_{\pm}}

\def\tk{\tilde{k}}

\def\Dp{\partial_+}
\def\Dm{\partial_-}
\def\Dpm{\partial_\pm}

\def\mDp{\mathcal{D}_{+}}
\def\mDm{\mathcal{D}_{-}}
\def\mDpm{\mathcal{D}_{\pm}}

\def\mFpm{\mathcal{F}_{+-}}

\def\adj{\textrm{adj}}

\def\etaL{\eta_{\L}}
\def\etaR{\eta_{\R}}

\def\bt{{\tilde{\b}}}
\def\tN{{\tilde{\mathcal{N}}}}
\def\tA{{\tilde{A}}}
\def\curlP{{\mathcal{P}_-}}

\def\smlr{{\scaleto{L,R}{4.5pt}}}

\def\smk{{\scaleto{k}{5pt}}}
\def\smtk{{\scaleto{\tilde{k}}{5.5pt}}}

\def\smtb{{\scaleto{\tilde{b}}{5.5pt} }}

\def\wztb{\mbox{\tiny WZ} , \smtb}

\def\wzk{\mbox{\tiny WZ} , \smk}

\def\wztk{\mbox{\tiny WZ} , \smtk}

\def\mfk{\mbox{\tiny WZ} , \smk}

\ifpublic

\else

\newcounter{comcompt}
\newcounter{piqle}
\newcounter{treqle}
\newcounter{boxq}
\fi

\usepackage{fancyhdr}
\ifpublic\else
\fi

\makeatletter
\let\@keywords\@empty
\let\@subject\@empty
\providecommand{\keywords}[1]{\gdef\@keywords{#1}}
\providecommand{\subject}[1]{\gdef\@subject{#1}}
\def\thetitle{\@title}
\def\theauthor{\@author}
\def\thesubject{\@subject}
\def\thedate{\@date}
\def\thekeywords{\@keywords}
\makeatother
\AtBeginDocument{
\hypersetup{pdftitle={\thetitle}}
\hypersetup{pdfauthor={\theauthor}}
\hypersetup{pdfsubject={\thesubject}}
\hypersetup{pdfkeywords={\thekeywords}}}

\ifworking
\newcommand{\work}[1]{{\begingroup \makeatletter\def\f@size{8}
#1 \endgroup}}
\else
\newcommand{\work}[1]{\ignorespaces}
\fi
\newcommand{\workstep}[1]{{\scriptsize $\bullet$~#1 \makeatletter\def\f@size{8}
}}

\title{Three-parameter integrable deformation of \texorpdfstring{\\}{}
\texorpdfstring{$\mathbb{Z}_4$}{Z4} permutation supercosets}
\author{F. Delduc, B. Hoare, T. Kameyama, S. Lacroix, M. Magro}

\begin{document}
\begin{flushright}
[ZMP-HH/18-22]
\end{flushright}
\begin{center}
\vspace*{2cm}

\begingroup\Large\bfseries\thetitle\par\endgroup

\vspace{1.5cm}

\begingroup
F. Delduc$^{a,}$\footnote{E-mail:~francois.delduc@ens-lyon.fr},
B. Hoare$^{b,}$\footnote{E-mail:~bhoare@itp.phys.ethz.ch},
T. Kameyama$^{a,}$\footnote{E-mail:~takashi.kameyama@ens-lyon.fr},
S. Lacroix$^{c,}$\footnote{E-mail:~sylvain.lacroix@desy.de},
M. Magro$^{a,}$\footnote{E-mail:~marc.magro@ens-lyon.fr}
\endgroup

\vspace{1cm}

\begingroup
$^a$\it Univ Lyon, Ens de Lyon, Univ Claude Bernard, CNRS, Laboratoire de Physique,
\\
F-69342 Lyon, France

\smallskip

$^b$ \it Institut f\"ur Theoretische Physik, Eidgen\"ossische Technische Hochschule Z\"urich,
\\
Wolfgang-Pauli-Strasse 27, 8093 Z\"urich, Switzerland

\smallskip

$^c$\it II. Institut f\"ur Theoretische Physik, Universit\"at Hamburg,
\\
Luruper Chaussee 149, 22761 Hamburg, Germany
\\
Zentrum f\"ur Mathematische Physik, Universit\"at Hamburg, \\
Bundesstrasse 55, 20146 Hamburg, Germany
\endgroup

\end{center}

\vspace{1cm}

\begin{abstract}
A three-parameter integrable deformation of $\mathbb{Z}_4$ permutation supercosets is constructed.
These supercosets are of the form $ {\widehat{F}} / {F_0}$ where $F_0$ is the bosonic diagonal subgroup of the product supergroup $\widehat{F}=\widehat{G} \times \widehat{G}$.
They include the AdS$_3$ $\times$ S$^3$ and AdS$_3$ $\times$ S$^3$ $\times$ S$^3$ supercosets.
This deformation encompasses both the bi-Yang-Baxter deformation of the semi-symmetric space $\sigma$-model on $\mathbb{Z}_4$ permutation supercosets and the mixed flux model.
Truncating the action at the bosonic level, we show that one recovers the bi-Yang-Baxter deformation of the principal chiral model plus Wess-Zumino term.
\end{abstract}

\newpage

\tableofcontents

%%%%%%%%%%%%%%%%%%%%%%%%%%%%%%%%%%%%%%%%%%%%%%%%%%%%%%%%%%%%%%%%%%%%%%%%%%%%%%%%
\section{Introduction}
\label{sec intro}

In this article we investigate integrable deformations of the semi-symmetric space $\sigma$-model on $\mathbb{Z}_4$ permutation supercosets.
These are supercosets that take the form
\begin{equation}\label{eq:supercoset}
\frac{\widehat{F}}{F_0} = \frac{\widehat{G} \times \widehat{G}}{F_0} ,
\end{equation}
where $F_0$ is the bosonic diagonal subgroup of the superisometry group $\widehat{G} \times \widehat{G}$.
The bosonic truncation of the undeformed theory yields the symmetric space $\sigma$-model on the $\mathbb{Z}_2$ permutation coset
\begin{equation}\label{eq:bosoniccoset}
\frac{F_0 \times F_0}{F_0} ,
\end{equation}
which is equivalent to the principal chiral model on $F_0$.
In \cite{Delduc:2017fib} a three-parameter integrable deformation of this bosonic model was constructed, generalising the $SU(2)$ case discussed in \cite{Lukyanov:2012zt}.
This deformed model can be understood as the bi-Yang-Baxter deformation of the principal chiral model plus Wess-Zumino term.
Our aim in this article is to construct the analogous three-parameter integrable deformation of the semi-symmetric space $\sigma$-model on the $\mathbb{Z}_4$ permutation supercoset \eqref{eq:supercoset}.

\medskip

The Yang-Baxter deformation of the principal chiral model was first introduced in \cite{Klimcik:2002zj}, generalising the one-parameter anisotropic $SU(2)$ principal chiral model of \cite{Cherednik:1981df} to general Lie groups.
The construction of a Lax pair for this model in \cite{Klimcik:2008eq} and the proof of Hamiltonian integrability in \cite{Delduc:2013fga} rely on the deformation being governed by a solution of the modified classical Yang-Baxter equation.
The two-parameter bi-Yang-Baxter deformation of the principal chiral model was introduced in \cite{Klimcik:2008eq} and shown to be integrable in \cite{Klimcik:2014bta,Delduc:2015xdm}.
As demonstrated in \cite{Hoare:2014pna}, the bi-Yang-Baxter deformation of the $SU(2)$ principal chiral model is equivalent to the two-parameter deformation of S$^3$ found in \cite{Fateev:1996ea}.
In \cite{Delduc:2013fga} the Yang-Baxter deformation of the symmetric space $\sigma$-model was constructed and shown to be integrable.
For symmetric spaces that take the form of $\mathbb{Z}_2$ permutation cosets \eqref{eq:bosoniccoset}, the deformed model is equivalent to a particular one-parameter model contained within the bi-Yang-Baxter deformation of the principal chiral model on $F_0$.

Adding the standard topological Wess-Zumino term \cite{Novikov:1982ei,Witten:1983tw,Witten:1983ar} to the principal chiral model is well-known to preserve its classical integrability.
Up to the expected quantization of the level, the resulting model interpolates between the principal chiral and conformal Wess-Zumino-Witten models.
Generalising the $SU(2)$ construction of \cite{Kawaguchi:2011mz,Kawaguchi:2013gma}, the Yang-Baxter deformation of the principal chiral model plus Wess-Zumino term for general Lie groups was derived in \cite{Delduc:2014uaa}.
In order to construct this deformation it was assumed that the solution of the modified classical Yang-Baxter deformation governing the deformation cubes to its negative.
This assumption is natural since it is satisfied by the standard Drinfel'd-Jimbo $\mathcal{R}$-matrix for a non-split real form of a semi-simple Lie algebra \cite{Drinfeld:1985rx,Jimbo:1985zk,Belavin:1984}.
In this article we will continue to focus on deformations governed by such $\mathcal{R}$-matrices.

By allowing for an antisymmetric term in the action, the two-parameter deformation of \cite{Fateev:1996ea} was generalised to an integrable four-parameter deformation of S$^3$ in \cite{Lukyanov:2012zt}.
Observing that this model also contains the TsT transformation of the $SU(2)$ Wess-Zumino-Witten model \cite{Horne:1991gn,Giveon:1991jj}, it was proposed in \cite{Hoare:2014pna} that the four-parameter model should be understood as the combined bi-Yang-Baxter deformation and TsT transformation of the $\sigma$-model on S$^3$ plus Wess-Zumino term.
The bi-Yang-Baxter deformation of the principal chiral model plus Wess-Zumino term was constructed in \cite{Delduc:2017fib}.
Applying a TsT transformation in the directions of the Cartan subalgebra, which is associated with adding a compatible abelian solution of the classical Yang-Baxter equation \cite{Kawaguchi:2014qwa,Matsumoto:2014nra,Matsumoto:2014gwa,Matsumoto:2015jja,Matsumoto:2015uja,vanTongeren:2015soa,Vicedo:2015pna,Osten:2016dvf} to the Drinfel'd-Jimbo $\mathcal{R}$-matrix, and considering the $SU(2)$ case it was shown that this model indeed generalises the four-parameter deformation of \cite{Lukyanov:2012zt} to general Lie groups.

\medskip

The semi-symmetric space $\sigma$-model of \cite{Metsaev:1998it,Berkovits:1999zq} is integrable \cite{Bena:2003wd,Magro:2008dv,Vicedo:2010qd} and describes the Green-Schwarz superstring \cite{Green:1983wt,Grisaru:1985fv,Tseytlin:1996hs,Cvetic:1999zs,Wulff:2013kga}, and consistent truncations thereof, on various supergravity backgrounds supported by Ramond-Ramond flux \cite{Zarembo:2010sg}.
Included amongst these are the AdS$_3 \times$ S$^3 \times$ T$^4$ \cite{Pesando:1998wm,Rahmfeld:1998zn,Park:1998un,Chen:2005uj} and AdS$_3 \times$ S$^3 \times$ S$^3 \times$ S$^1$ \cite{Babichenko:2009dk} backgrounds, each preserving 16 supersymmetries, which, in the context of this article, are of particular interest as the corresponding $\sigma$-models are on $\mathbb{Z}_4$ permutation supercosets \eqref{eq:supercoset} with $\widehat{F} = PSU(1,1|2)$ and $\widehat{F} = D(2,1;\alpha)$ respectively.

Furthermore, it is well-known that both these two backgrounds can be supported by either Ramond-Ramond (R-R) flux, Neveu-Schwarz-Neveu-Schwarz (NS-NS) flux or a combination of the two.
Classically, there is a one-parameter family of backgrounds interpolating between the pure R-R and pure NS-NS cases, whose bosonic truncation is the principal chiral model plus Wess-Zumino term.
The corresponding one-parameter deformation of the semi-symmetric space $\sigma$-model, the mixed flux model, together with the Lax pair demonstrating its integrability, was derived in \cite{Cagnazzo:2012se}.

In \cite{Delduc:2013qra,Delduc:2014kha} the Yang-Baxter deformation of the semi-symmetric space $\sigma$-model for general $\mathbb{Z}_4$ supercosets was constructed.
In the case the supercoset takes the form of a $\mathbb{Z}_4$ permutation supercoset \eqref{eq:supercoset} it is natural to expect that the model admits a two-parameter bi-Yang-Baxter deformation.
This was indeed shown to be the case in \cite{Hoare:2014oua}.
In this article we generalise these constructions to find the bi-Yang-Baxter deformation of the mixed flux model of \cite{Cagnazzo:2012se} giving a three-parameter integrable deformation of the semi-symmetric space $\sigma$-model on the $\mathbb{Z}_4$ permutation supercosets \eqref{eq:supercoset}.

\medskip

The plan of this article is as follows.
In section \ref{sec: summary} we present the action of the three-parameter deformation in two forms.
The first involves additional gauge and auxiliary fields, the inclusion of which clarifies the underlying structure of the model.
The additional fields are therefore particularly useful for constructing the Lax pair and demonstrating the classical integrability of the model, which is discussed in section \ref{sec: eom mce}.
The second form is the action that follows from integrating out the additional fields, the explicit computation of which we give in section \ref{sec: without A}.
In section \ref{sec: truncation limits} we demonstrate agreement with various known truncations and limits.
First, in subsection \ref{subsec: truncation}, we consider the bosonic truncation and relate it to the model of \cite{Delduc:2017fib}.
We then study the limits that give the mixed flux model of \cite{Cagnazzo:2012se} and the two-parameter bi-Yang-Baxter deformation of \cite{Hoare:2014oua} in subsections \ref{subsec: mf} and \ref{subsec: sbyb} respectively.
We conclude with some possible future directions and applications in section \ref{sec: conclusion}.
There are three appendices.
In appendix \ref{app: coeffs} we give the explicit form of coefficients of the linear maps used in section \ref{sec: eom mce} to construct the Lax pair.
In appendix \ref{apps: dhmt} we start from the action of \cite{Delduc:2017fib} and rewrite it in a form suitable for comparing with the bosonic truncation obtained in subsection \ref{subsec: truncation}.
Finally, in appendix \ref{app:mbads3s3}, we rewrite the metric and B-field of the
three-parameter deformation of S$^3$, first given in \cite{Lukyanov:2012zt}, in
a particularly simple form.

%%%%%%%%%%%%%%%%%%%%%%%%%%%%%%%%%%%%%%%%%%%%%%%%%%%%%%%%%%%%%%%%%%%%%%%%%%%%%%%%
\section{Action of the three-parameter deformation}
\label{sec: summary}

In this section, we give the definition of the three-parameter integrable deformation of $\mathbb{Z}_4$ permutation supercosets.
As stated in the introduction and proved in subsection \ref{subsec: truncation}, the bosonic truncation of the integrable $\sigma$-model presented here is in agreement with the three-parameter integrable deformation of $\mathbb{Z}_2$ permutation cosets built in \cite{Delduc:2017fib}.
This model generalised the Yang-Baxter deformation of the principal chiral model plus Wess-Zumino term constructed in \cite{Delduc:2014uaa}.
To construct an integrable deformation one needs to demonstrate the existence of a Lax pair.
On a technical level, this requires the inversion of some operator involving the $\mathcal{R}$-matrix, a skew-symmetric solution of the modified classical Yang-Baxter equation (mCYBE) used to define the Yang-Baxter deformation.
As shown in \cite{Delduc:2014uaa}, for the Yang-Baxter deformation of the principal chiral model plus Wess-Zumino term this inversion is tractable when the $\mathcal{R}$-matrix takes the standard form.

To combine the bi-Yang-Baxter deformation and the WZ term the strategy developed in \cite{Delduc:2017fib} is based on formulating the principal chiral model as a $\mathbb{Z}_2$ permutation coset \eqref{eq:bosoniccoset}.
As shown in \cite{Delduc:2017fib} introducing a gauge field makes the inversion of the relevant operators, which is necessary to prove the existence of a Lax pair, tractable.
To go further and construct the action in the supercoset case we also introduce auxiliary fields.
In this section we therefore give two formulations of the model.
The first corresponds to the action including the gauge and auxiliary fields.
It is this formulation that enables us to construct a Lax pair.
The second is obtained after eliminating the gauge and auxiliary fields.

\paragraph{Algebraic setting and related definitions.}

We consider supercosets of the type $ {\widehat{F}} / {F_0}$ where $F_0$ is the bosonic diagonal subgroup of the product supergroup $\widehat{F}=\widehat{G} \times \widehat{G}$.
The superalgebras $\hat{\mathfrak{g}}$ and $\hat{\mathfrak{f}}=\hat{\mathfrak{g}} \oplus \hat{\mathfrak{g}}$ correspond to the supergroups $\widehat{G}$ and $\widehat{F}$ respectively.
We shall make use of the standard block-diagonal matrix realisation of the product supergroup $\widehat{F}$.
At the level of the superalgebra, for any element $\mathcal{X} = \textrm{diag} ( \XL , \XR ) \in \hat{\mathfrak{f}}$ with $ \XL , \XR \in \hat{\mg}$, we define the supertrace of ${\mathcal{X}}$ as $\STr({\mathcal X})= \STr({\XL})+\STr({\XR})$.
The superalgebra $\hat{\mathfrak{f}}$ has a $\mathbb{Z}_4$ decomposition:
\beqz
\hat{\mathfrak{f}} = {\mathfrak{f}}_0 \oplus {\mathfrak{f}}_1 \oplus {\mathfrak{f}}_2
\oplus {\mathfrak{f}}_3,
\eeqz
with ${\mathfrak{f}}_0$ the Lie algebra associated with the Lie group $F_0$.
Let us denote $P_B$ and $P_F$ the projections onto the even and odd parts of $\hat{\mg}$.
The $\mathbb{Z}_4$ decomposition of $\mathcal{X}$ is defined as \cite{Babichenko:2009dk}
\bse \label{eq: z4grading}
\begin{align}
P_0 \mathcal{X} &= \mathcal{X}^{(0)} =\tfrac12 \begin{pmatrix}
P_B ( \XL + \XR ) & 0 \\
0 & P_B ( \XL + \XR ) \\
\end{pmatrix} , \\
P_1 \mathcal{X} &= \mathcal{X}^{(1)} =\tfrac12 \begin{pmatrix}
P_F ( \XL + i \XR ) & 0 \\
0 & -i P_F ( \XL + i \XR ) \\
\end{pmatrix} , \\
P_2 \mathcal{X} &= \mathcal{X}^{(2)} =\tfrac12 \begin{pmatrix}
P_B ( \XL - \XR ) & 0 \\
0 & -P_B ( \XL - \XR ) \\
\end{pmatrix} , \\
P_3 \mathcal{X} &= \mathcal{X}^{(3)} =\tfrac12 \begin{pmatrix}
P_F ( \XL - i \XR ) & 0 \\
0 & i P_F ( \XL - i \XR ) \\
\end{pmatrix}.
\end{align}
\ese
This decomposition is such that for any ${\mathcal X}, \mathcal{Y} \in \hat{\mathfrak{f}}$ we have $\STr({\mathcal{X}}^{(a)} \, {\mathcal{Y}}^{(b)}) =0$ if $a+b \neq 0$~mod~4.
Finally, we introduce the matrix $W = \textrm{diag} (1, -1)$, which satisfies the relations
\beqz
P_0 W = W P_2 , \qquad P_2 W = W P_0 , \qquad P_1 W = W P_3 , \qquad P_3 W = W P_1 .
\eeqz

\paragraph{Action with gauge and auxiliary fields.}

The action depends on three parameters $\ELR$ and $k$.
The dynamical field of the model is
\beqz
g = \textrm{diag}( \gL, \gR) \quad \mbox{with} \quad \gLR(\sigma^\pm) \in \widehat{G},
\eeqz
and where $\sigma^\pm$ are light-cone coordinates.
The left-invariant current is defined as
\beqz
\mJpm = g^{-1} \partial_\pm g = \textrm{diag}( \gL^{-1} \partial_\pm \gL,
\gR^{-1} \partial_\pm \gR) ,
\eeqz
with $\partial_\pm = \partial_0 \pm \partial_1$.
We also introduce $\mApm$ taking values in $\hat{\mathfrak{f}}$.
As we shall see, the grade zero part, $\mApm^{(0)}$, plays the role of a gauge field while the other gradings, $\mApm^{(a)}$ with $a \neq 0$, are auxiliary fields.
The definition of the action with gauge and auxiliary fields is
\beq\begin{split}
S_{\eta_\smlr , \smk} [g, \mApm] &= \int d^2\sigma \, \, \STr
\Bigl[\left( \mJp -\mAp \right) \mOm \left( \mJm - \mAm \right)
+ \mAp \bigl( \hTm + k W \left( 1 + \hTm \right) \bigr) \mAm \Bigr]\\
& \qquad + k \, \STr \Bigl[
W \bigl( \mJp \mAm -   \mAp \mJm
+
\mJm^{(2)} \mJp^{(0)} - \mJm^{(0)}  \mJp^{(2)}
\bigr)
\Bigr] + S_{\mfk} [g] ,
\label{L0}
\end{split}\eeq
where the non-standard Wess-Zumino (WZ) term of \cite{Cagnazzo:2012se}, in the form already used in \cite{Hoare:2013pma,Babichenko:2014yaa,Hoare:2014oua}, is given by
\beq
S_{\mfk} [g] = - 4 k \int d^2\sigma d\xi \, \, \epsilon^{\mu \nu \rho}
\STr
\Bigl[ \tfrac23 W \mJ_\mu^{(2)} \mJ_\nu^{(2)} \mJ_\rho^{(2)} +
W [ \mJ_\mu^{(1)}, \mJ_\nu^{(3)} ] \mJ_\rho^{(2)}
\Bigr],
\label{LWZ}
\eeq
with $ \epsilon^{\mu \nu \rho}$ completely antisymmetric.

The operator $\hTm$ is defined as the following linear combination of projectors,
\beq
\hTm = - 2 P_2 + \frac{ \lambda P_1 - k W P_3}{1 - \lambda} -
\frac{\lambda P_3 + k W P_1}{1 + \lambda},
\label{hTm}
\eeq
with
\beq
\lambda = \sqrt{ \frac{( 1 - k^2 -\EL^2 ) ( 1 - k^2 -\ER^2 )}{ 1 - k^2 } } .\label{ansatz lambda}
\eeq

\medskip

As usual, we introduce a standard skew-symmetric solution $\mR$ of the
modified classical Yang-Baxter equation on $\hat{\mathfrak{f}}$.
It therefore satisfies the three properties
\begin{equation}\begin{gathered} \label{eq2.8}
[ \mR \mX , \mR \mY ] = \mR \left( [ \mR \mX , \mY ] + [ \mX , \mR \mY ] \right) + [ \mX , \mY ] ,
\\
\STr [ \mX \, \mR \mY ] = - \STr [ \mR \mX \, \mY ] , \qquad \qquad \mR^3 = - \mR ,
\end{gathered}\end{equation}
for any $\mX, \mY \in \hat{\mathfrak{f}}$. Since $\mR$ is a standard
solution of the mCYBE, the operator
\beqz
\Pi = 1 + \mR^2
\eeqz
is the projector onto the Cartan subalgebra of $\hat{\mathfrak{f}}$ and satisfies
(see for instance \cite{Delduc:2014uaa})
\begin{equation}
\Pi \mR = \mR \Pi = 0, \qquad \qquad
\Pi[\mR\mX,\mY] + \Pi[\mX,\mR\mY] = 0 , \qquad \mX,\mY \in \hat{\mathfrak{f}}. \label{eq:pirbracket}
\end{equation}
As a consequence of the property $\mR^3 = - \mR$ we also have the following simple inversion formula
\begin{equation}
( \Pi + \beta \mR + \gamma \mR^2 )^{-1} = \Pi + \frac{1}{\beta^2 + \gamma^2}
( - \beta \mR + \gamma \mR^2).
\label{eq:formulainversionR}
\end{equation}

The operator $\mOm$ is defined in terms of the dressed operators
\beqz
\mRg = \Ad_g^{-1} \circ \mR \circ \Ad_g \quad \mbox{and} \quad
\Pi_g = \Ad_g^{-1} \circ \Pi \circ \Ad_g,
\eeqz
such that the operator $\mRg$ is also a skew-symmetric solution of the mCYBE. We then introduce
the operators
\beq \label{Omgpm}
\Omega_\pm = \pm \frac{\sqrt{(\mu - 1) (1 - k^2 \mu)}}{1 \mp k W} \,
\mRg \pm k W \left(\frac{\mu - 1}{1 \mp k W} \right) \mRg^2.
\eeq
In this expression, $\mu$ is a block-diagonal matrix depending
on the three parameters $\ELR$ and $k$:
\beq
\mu = 1 + \frac{1}{\lambda^2+k^2} \, \text{diag} (\EL^2 , \ER^2).
\label{ansatz}
\eeq
Finally, the operator $\mOm$ appearing in the action \eqref{L0} is given by
\beq
\mOpm = \frac{1 \pm k W \Omega_\pm}{1 + \Omega_\pm},
\label{def mOpm}
\eeq
where for later use we have also introduced $\mOp = (\mOm)^t$.

\paragraph{Action after elimination of $\mApm$.}

In section \ref{sec: without A} we compute the action
obtained after elimination of the gauge and
auxiliary fields $\mApm$. The result is given in equation \eqref{Lfullresult2}. For convenience,
we reproduce it here. It takes the form
\beq
S_{\eta_\smlr , \smk} [g] =
\int d^2 \sigma \, \,
\STr\Bigl[\mJp \, \mSm \mJm\Bigr] + S_{\mfk} [g]. \label{Lfullresulttext}
\eeq
The definition of the operator $\mSm$ is
\beq
\mSm = \Bigl ( \left( 1 - \left( 2 P_2 + P_F \right) k W \Omega_- \right) d_-
+ P_F \, k W \left( 1 - d_- \right) \Bigr ) \left( 1 + \Omega_- d_- \right)^{-1} ,
\label{mSmresult}
\eeq
with
\beq \label{eq def dm}
\begin{split}
d_- =
2 P_2 &+ \frac{1}{1 - k^2} \Bigl( \left( \lambda - k^2 \right) P_1 - \left( 1 + \lambda \right) k W P_3 \Bigr)
\\
&- \frac{1}{1 - k^2} \Bigl(\left( \lambda + k^2 \right) P_3 + \left( 1 - \lambda \right) k W P_1 \Bigr) .
\end{split}
\eeq

\paragraph{Gauge invariance.}

As we shall indicate in section \ref{sec: eom mce} for the action
\eqref{L0} and prove in
subsection \ref{subsec: proof ginv} for the action \eqref{Lfullresulttext}, the field theory
constructed in this article
is on the supercoset $ {\widehat{F}} / {F_0}$. This is the consequence
of the existence of a gauge invariance under $F_0$. More precisely, the
corresponding gauge transformations are
\beq
g \to g g_0, \qquad
\mApm \to g_0^{-1} \partial_\pm g_0 + g_0^{-1} \mApm g_0, \label{gauge transfos}
\eeq
with $g_0(\sigma^\pm)$ taking values in $F_0$. In particular, this means
that $\mApm^{(0)}$ transforms as
\beqz
\mApm^{(0)} \to g_0^{-1} \partial_\pm g_0 + g_0^{-1} \mApm^{(0)} g_0.
\eeqz
This shows that $\mApm^{(0)}$ is a gauge field. The other gradings $\mApm^{(a)}$ of
$\mApm$ with $ a \neq 0$, have the homogeneous gauge transformations
\beqz
\mApm^{(a)} \to g_0^{-1} \mApm^{(a)} g_0.
\eeqz
Therefore, they correspond to auxiliary fields.

%%%%%%%%%%%%%%%%%%%%%%%%%%%%%%%%%%%%%%%%%%%%%%%%%%%%%%%%%%%%%%%%%%%%%%%%%%%%%%%%
\section{Equations of motion, the Maurer-Cartan equation and the Lax pair}
\label{sec: eom mce}

In this section we will demonstrate the classical integrability of the bi-Yang-Baxter deformation
of the mixed flux model as defined
in section \ref{sec: summary}.
To do so we first compute the equations of motion following from the action \eqref{L0}.
These are of two types: two constraint equations arising from the variation with respect to $\mApm$ and a dynamical equation that comes from varying with respect to the supergroup-valued field $g$.
Working on the constraint equations, we show that the dynamical equation in first-order form, i.e. in terms of currents, and the Maurer-Cartan equation for the currents follow from the
zero curvature of a Lax pair.

\paragraph{Equations of motion for $\mApm$.}
Varying the action \eqref{L0} with respect to $\mAmp$, we find the constraint equations
\beq
( \mOpm \mp k W ) (\mJpm - \mApm) = \left( 1 \mp k W \right) \hTpm \mApm, \label{eom A}
\eeq
where we have introduced a second sum of projectors, $\hTp$, defined as
\beqz
\hTp = ( 1 - k W )^{-1} \Big( \big( \hTm + k W ( 1 + \hTm ) \big)^t + k W \Big) .
\eeqz
The linear combinations of projectors $\hTpm$ are given by
\beq
\hTpm = - 2 P_2 \mp \frac{ \lambda P_1 - k W P_3}{1\pm \lambda}
\pm \frac{\lambda P_3 + k W P_1}{1\mp\lambda} , \label{hTp}
\eeq
and satisfy the following relation
\beqz
\hTm{}^t = \hTp + \frac{4k}{\lambda^2-1} W ( P_1 + P_3) .
\eeqz
In order to show the existence of a Lax pair, it will be convenient to introduce
a new current $\mQpm$ defined as
\beq
\mQpm = (1+\Omega_\pm)^{-1} (\mJpm - \mApm ) , \label{def Qpm}
\eeq
where we recall that $\Omega_\pm$ are defined in \eqref{Omgpm}. In terms of this new
current, the equations of motion for $\mAmp$ \eqref{eom A} take the particularly simple form
\beq
\mQpm = \hTpm \mApm ,\label{eom A2}
\eeq
from which it follows that
\beqz
\mQpm^{(0)} = 0.
\eeqz

\paragraph{Equation of motion for $g$.}

Varying the action \eqref{L0} with respect to $g$ and eliminating $\mJpm$ in favour of $\mQpm$ \eqref{def Qpm} we find the following equation of motion
\beq
\mE \equiv \mDp (1+k W) \mQm + \mDm (1-kW) \mQp + 2 k W \mFpm(\mApm) = 0, \label{E}
\eeq
where
\beqz
\mFpm(\mApm) = \Dp \mAm - \Dm \mAp + [\mAp,\mAm] , \qquad \mDpm = \Dpm + \adj_{\mApm} ,
\eeqz
with $\adj_{\mApm} = [\mApm,\cdot]$.
Decomposing $\mE$ under the $\mathbb{Z}_4$ grading \eqref{eq: z4grading} we find that the grade 0 part is given by
\begin{equation*}\begin{split}
\mE^{(0)} = \, & kW (\Dp+\adj_{\mAp^{(0)}}) (\mQm^{(2)} + 2 \mAm^{(2)}) - kW (\Dm+\adj_{\mAm^{(0)}})(\mQp^{(2)} + 2 \mAp^{(2)})
\\ & +[\mAp^{(2)},\mQm^{(2)}] +[\mAm^{(2)},\mQp^{(2)}] + 2kW\big( [\mAp^{(1)},\mAm^{(1)}] + [\mAp^{(3)},\mAm^{(3)}]\big)
\\ & + [\mAp^{(1)},\mQm^{(3)}] + [\mAp^{(3)},\mQm^{(1)}] + [\mAm^{(1)},\mQp^{(3)}] + [\mAm^{(3)},\mQp^{(1)}]
\\ & + kW \big( [\mAp^{(1)},\mQm^{(1)}] + [\mAp^{(3)},\mQm^{(3)}] - [\mAm^{(1)},\mQp^{(1)}] - [\mAm^{(3)},\mQp^{(3)}] \big) .
\end{split}\end{equation*}
Evaluating on the constraint equations \eqref{eom A2} we find that $\mE^{(0)}$ identically vanishes.
This is a consequence of the gauge invariance \eqref{gauge transfos} of the action \eqref{L0}.

\paragraph{Maurer-Cartan equation.}
We now turn to the Maurer-Cartan equation
\beq
\Dp \mJm - \Dm \mJp + [ \mJp , \mJm ] = 0 , \label{zce}
\eeq
which we rewrite in terms of $\mQpm$ and $\mApm$ using
\beqz
\mJpm = \mApm + (1+\Omega_\pm) \mQpm ,
\eeqz
which follows from \eqref{def Qpm}.
As outlined in section \ref{sec: summary}, in equation \eqref{Omgpm} the operators $\Omega_\pm$ are defined in terms of a standard skew-symmetric solution $\mR$ of the mCYBE on $\hat{\mathfrak{f}}$.
In particular, we have
\beq\label{eq:omegapm}
1+\Omega_\pm = \Pi_g + \beta_\pm \mR_g + \gamma_\pm \mR_g^2 ,
\eeq
with the block diagonal matrix coefficients $\beta_\pm$ and $\gamma_\pm$ given by
\beq\label{eq:bd}
\beta_\pm = \pm \frac{\sqrt{(\mu-1)(1-k^2\mu)}}{1 \mp kW}, \qquad
\gamma_\pm = - \frac{1 \mp k\mu W}{1 \mp kW} .
\eeq
The Maurer-Cartan equation \eqref{zce} is then given by
\begin{equation}\begin{split}\label{eq:eq1}
\mFpm(\mApm) + (1+\Omega_-)\mDp\mQm - (1+\Omega_+)\mDm\mQp
- [(1+\Omega_+) \mQp,(1+\Omega_-)\mQm] &
\\ + (1+\Omega_-) [ (1+\Omega_+) \mQp,\mQm]
+ (1+\Omega_+)[\mQp,(1+\Omega_-)\mQm] & = 0.
\end{split}\end{equation}
Using the equation of motion for $g$ \eqref{E} to replace
\beqz
\mDp\mQm + \mDm\mQp \to - kW (\mDp\mQm - \mDm\mQp) - 2kW \mFpm(\mApm) ,
\eeqz
we substitute in the explicit expressions for $1+\Omega_\pm$ \eqref{eq:omegapm} and
simplify using the properties \eqref{eq2.8} and \eqref{eq:pirbracket}, to rewrite the Maurer-Cartan equation as
\begin{equation}\begin{split}\label{eq:eq2}
& O_1\mFpm(\mApm) + O_2(\mDp\mQm - \mDm\mQp) + O_3 [\mQp,\mQm]
\\
&-\beta_+(1+\gamma_-) \Pi_g[\mQp,\mR_g\mQm]-\beta_-(1+\gamma_+)\Pi_g[\mR_g\mQp,\mQm] = 0,
\end{split}\end{equation}
where
\begin{equation}\begin{split}\label{eq:operators}
O_1 & = \Pi_g + kW(\beta_+ - \beta_-) \mR_g + (kW(\gamma_+ - \gamma_-)-1) \mR_g^2 ,
\\
O_2 & = \Pi_g + \tfrac12(\beta_+ + \beta_- +kW(\beta_+-\beta_-))
\mR_g + \tfrac12(\gamma_+ + \gamma_-+kW(\gamma_+-\gamma_-)) \mR_g^2 ,
\\
O_3 & = - (\beta_+\beta_-+\gamma_+\gamma_- + \gamma_+ + \gamma_-)\Pi_g -
(\beta_+ \gamma_- + \gamma_+ \beta_-) \mR_g + (\beta_+ \beta_- - \gamma_+ \gamma_-)
\mR_g^2 .
\end{split}\end{equation}
From the definitions of $\beta_\pm$ and $\gamma_\pm$ \eqref{eq:bd} one can check that
\beq \label{eqtobesatisfied}
\beta_+ (1+ \gamma_-) = \beta_- (1+ \gamma_+) .
\eeq
This, together with the identity \eqref{eq:pirbracket},
implies that the left-hand side of the second line of \eqref{eq:eq2} is identically zero.
Furthermore, again using the definitions of $\beta_\pm$ and $\gamma_\pm$,
the three operators \eqref{eq:operators} can be seen to be block proportional to each other with
\beqz
O_1 = O_2 = \mu^{-1} O_3 .
\eeqz
Together, this brings us to our final form of the Maurer-Cartan equation
\beq\label{Z}
\mZ \equiv \mDp\mQm - \mDm\mQp + \mu [\mQp,\mQm] + \mFpm(\mApm) = 0 .
\eeq

\paragraph{Lax pair.}

In order to construct a Lax pair we work on the constraint equations \eqref{eom A2} and understand the equation of motion \eqref{E} and the Maurer-Cartan equation \eqref{Z} as a set of two first-order equations for $\mathcal{A}_\pm$.
We then attempt to construct two linear maps
\begin{equation} \label{eq:linmapA}
\mathcal{A}_\pm = \sum_{i=0}^3 \big( b_{\pm}^i P_i \mathcal{K}_\pm + c_{\pm}^{i} W P_i \mathcal{K}_\pm \big) ,
\end{equation}
and (recalling that $\mathcal{E}^\grade{0} = 0$) the two combinations
\begin{equation} \label{eq:eezz}\begin{aligned}
\widetilde{\mathcal{E}} & = \sum_{i=0}^3 \big( e_i P_i \mathcal{E} + \tilde{e}_i W P_i \mathcal{E} + w_i P_i \mathcal{Z} + \tilde{w}_i W P_{i} \mathcal{Z} \big),
\\
\widetilde{\mathcal{Z}} & = \sum_{i=0}^3 \big( z_i P_i \mathcal{Z} + \tilde{z}_i W P_i \mathcal{Z} + f_i P_i \mathcal{E} + \tilde{f}_i W P_{i} \mathcal{E} \big),
\end{aligned}\end{equation}
such that
\begin{equation}\begin{aligned}\label{eq:finaleq}
\widetilde{\mathcal{E}}^\grade{2} & = D^\grade{0}_+ \mathcal{K}_-^\grade{2} + D^\grade{0}_- \mathcal{K}_+^\grade{2} + [\mathcal{K}_+^\grade{1},\mathcal{K}_-^\grade{1}] - [\mathcal{K}_+^\grade{3},\mathcal{K}_-^\grade{3}],
\\
\widetilde{\mathcal{E}}^\grade{0} & = 0, \qquad
\widetilde{\mathcal{E}}^\grade{1} = [\mathcal{K}_+^\grade{2} ,\mathcal{K}_-^\grade{3}] ,\qquad
\widetilde{\mathcal{E}}^\grade{3} = [\mathcal{K}_+^\grade{1} ,\mathcal{K}_-^\grade{2}] ,
\\
\widetilde{\mathcal{Z}}^\grade{0} & = F_{+-}^\grade{0} + [\mathcal{K}_+^\grade{2},\mathcal{K}_-^\grade{2}] + [\mathcal{K}_+^\grade{1},\mathcal{K}_-^\grade{3}] + [\mathcal{K}_+^\grade{3},\mathcal{K}_-^\grade{1}] ,
\\
\widetilde{\mathcal{Z}}^\grade{2} & = D^\grade{0}_+ \mathcal{K}_-^\grade{2} - D^\grade{0}_- \mathcal{K}_+^\grade{2} + [\mathcal{K}_+^\grade{1},\mathcal{K}_-^\grade{1}] + [\mathcal{K}_+^\grade{3},\mathcal{K}_-^\grade{3}],
\\
\widetilde{\mathcal{Z}}^\grade{1} & = D_+^\grade{0} \mathcal{K}_-^\grade{1} - D_-^\grade{0}\mathcal{K}_+^\grade{1} + [\mathcal{K}_+^\grade{2},\mathcal{K}_-^\grade{3}] + [\mathcal{K}_+^\grade{3},\mathcal{K}_-^\grade{2}] ,
\\
\widetilde{\mathcal{Z}}^\grade{3} & = D_+^\grade{0} \mathcal{K}_-^\grade{3} - D_-^\grade{0}\mathcal{K}_+^\grade{3} + [\mathcal{K}_+^\grade{2},\mathcal{K}_-^\grade{1}] + [\mathcal{K}_+^\grade{1},\mathcal{K}_-^\grade{2}] ,
\end{aligned}\end{equation}
where
\beqz
D^\grade{0}_\pm = \partial_\pm + \adj_{\mathcal{K}^\grade{0}_\pm} , \qquad
F^\grade{0}_{+-} = \partial_+ \mathcal{K}^\grade{0}_- - \partial_- \mathcal{K}^\grade{0}_+
+ [\mathcal{K}^\grade{0}_+ , \mathcal{K}^\grade{0}_-] .
\eeqz
Such linear maps can indeed be found, with $b_\pm^i$ and $c_\pm^i$, along with the remaining parameters, given in
appendix \ref{app: coeffs}. The equations
\beqz
\widetilde{\mathcal{E}}^\grade{i} = \widetilde{\mathcal{Z}}^\grade{i} = 0,
\eeqz
as defined in \eqref{eq:finaleq} then take the form of the familiar first-order
equations of the semi-symmetric space $\sigma$-model \cite{Metsaev:1998it}.
It immediately follows that the Lax pair for the three-parameter deformation
is given by \cite{Bena:2003wd}
\beq \label{Lax}
\mL_\pm (z) = \mKpm^{(0)} + z^{-1} \, \mKpm^{(1)} + z^{\pm 2} \, \mKpm^{(2)} + z \, \mKpm^{(3)} ,
\eeq
where $\mKpm$ are defined in terms of $\mJpm$ through
\eqref{eq:linmapA}, \eqref{eom A2} and \eqref{def Qpm}.

\paragraph{Comment.} At this point we should emphasise that the existence of the Lax pair is highly non-trivial.
Its existence stems from the peculiar form of the action \eqref{L0}.
This includes, in particular, the very specific way in which the auxiliary fields appear as well as the tuning of the many coefficients that enter into its definition.
Let us therefore stress that the action \eqref{L0} does not come from nowhere!
It is the result of a thorough and rather complicated investigation.

%%%%%%%%%%%%%%%%%%%%%%%%%%%%%%%%%%%%%%%%%%%%%%%%%%%%%%%%%%%%%%%%%%%%%%%%%%%%%%%%
\section{Elimination of gauge and auxiliary fields}
\label{sec: without A}

The reason for introducing the gauge and auxiliary fields is that it enabled us to determine the Lax pair as demonstrated in the previous section.
We now compute the action obtained after elimination of the gauge and auxiliary fields $\mApm$.

\paragraph{Equations of motion for $\mApm$.}

We first determine the on-shell values of $\mApm$ in terms of $\mJpm$.
For this we combine \eqref{def Qpm} and \eqref{eom A2} to obtain
\beq
\mJpm - \mApm = ( 1 + \Omega_\pm ) \hTpm \mApm.
\label{J-A}
\eeq
Defining
\beqz
d_\pm = \frac{\hTpm}{1 + \hTpm},
\eeqz
equation \eqref{J-A} may be rewritten as
\beqz
\mJpm - \mApm = \hTpm \mApm + \Omega_\pm d_\pm (1 + \hTpm) \mApm.
\eeqz
We therefore have
\beq
\mApm
= \left( 1 + \hTpm \right)^{-1} \left( 1 +
\Omega_\pm d_\pm \right)^{-1} \mJpm,
\label{sol A}
\eeq
which are the on-shell expressions for $\mApm$.

\paragraph{Action.}

One way to proceed is to rewrite the action \eqref{L0} as
\begin{equation}\begin{split}
S_{\eta_\smlr , \smk} [g, \mApm] = - & \int d^2\sigma \, \,
\STr \Bigl[
\mAp \bigl( \left( \mOm + k W \right) \left( \mJm - \mAm \right) - ( 1 + k W ) \hTm \mAm \bigr)
+ \mJp {\mathcal{C}}_-
\Bigr]
\\ & 
+ S_{\mfk} [g], \label{L1}
\end{split}\end{equation}
where
\beq
{\mathcal{C}}_- = \left( \mOm + k W \right) \left( \mJm - \mAm \right) + k W \left( 2 P_2 + P_F \right) \mAm
- k W \left( 2 P_0 + P_F \right) \left( \mJm - \mAm \right) .
\eeq
The first term in the action \eqref{L1} vanishes upon imposing the equation of motion \eqref{eom A} for $\mAp$.
It remains therefore to compute ${\mathcal{C}}_-$ on-shell. We find
\bea
{\mathcal{C}}_- &=& ( 1 + k W ) \hTm \mAm + k W \left( 2 P_2 + P_F \right) \mAm - k W \left( 2 P_0
+ P_F \right) \left( 1 + \Omega_- \right) \hTm \mAm \label{eq:preshow} \\
&=& \bigl( \left( 1 - \left( 2 P_2 + P_F \right) k W \Omega_- \right) d_-
+ P_F \, k W \left( 1 - d_- \right) \bigl) \left( 1 + \hTm \right) \mAm, \label{show}
\eea
where we have first used the relations \eqref{eom A} and \eqref{J-A} to arrive at \eqref{eq:preshow}.
We can now replace $\mAm$ in terms of $\mJm$ using its on-shell expression \eqref{sol A} to obtain
\beq
S_{\eta_\smlr,\smk}[g] =
\int d^2 \sigma \, \,
\STr\Bigl[\mJp \, \mSm \mJm\Bigr] + S_{\mfk} [g], \label{Lfullresult2}
\eeq
where the operator $\mSm$ takes the form
\beq
\mSm = 2 \, \curlP \left( 1 + \Omega_- d_- \right)^{-1}, \label{mSmresult2}
\eeq
with the quantity $\curlP$ defined as
\beq
\curlP = \tfrac12 \Bigl( \left( 1 - \left( 2 P_2 + P_F \right) k W \Omega_- \right) d_-
+ P_F \, k W \left( 1 - d_- \right) \Bigr). \label{eq: defcurlP}
\eeq
The introduction of the operator $\curlP$ is useful to take the limits corresponding to both the mixed flux model and the bi-Yang-Baxter deformation in subsections \ref{subsec: mf} and \ref{subsec: sbyb} respectively.

\paragraph{Proof of gauge invariance.}
\label{subsec: proof ginv}

We can now present the postponed proof of the $F_0$-gauge invariance
of the field theory we have constructed. Under the gauge transformation
\eqref{gauge transfos}, the non-standard Wess-Zumino term $S_{\mfk} [g]$ is invariant.
The operator $\Omega_-$ defined by \eqref{Omgpm} transforms as
\beqz
\Omega_- \to g_0^{-1} \Omega_- g_0.
\eeqz
Due to the presence of the projectors $P_2$, $P_F$ and of
$d_-$, defined by \eqref{eq def dm}, in the expression of $\mSm$,
the only gradings of $\mJpm^{(a)}$ which contribute to the first term
of the action \eqref{Lfullresult2} are the non zero ones. As these
components of the currents have homogeneous gauge transformations,
the action \eqref{Lfullresult2} is gauge invariant.

%%%%%%%%%%%%%%%%%%%%%%%%%%%%%%%%%%%%%%%%%%%%%%%%%%%%%%%%%%%%%%%%%%%%%%%%%%%%%%%%
\section{Bosonic truncation and limits}
\label{sec: truncation limits}

In this section we demonstrate agreement with three known results corresponding to certain truncations and limits.
The first corresponds to the bosonic truncation, for which we recover the three-parameter deformation of $\mathbb{Z}_2$ permutation cosets worked out in \cite{Delduc:2017fib}.
The second and third cases are the mixed flux model and the bi-Yang-Baxter deformation respectively.
These are obtained when $\ELR=0$ and $k=0$ respectively.
The mixed flux model has been constructed in \cite{Cagnazzo:2012se} while the bi-Yang-Baxter deformation of $\mathbb{Z}_4$ permutation supercosets has been
found in \cite{Hoare:2014oua}.
The corresponding actions have been constructed without a gauge or auxiliary fields.
Therefore, here we make the comparison using the action \eqref{Lfullresult2}.

%%%%%%%%%%%%%%%%%%%%%%%%%%%%%%%%%%%%%%%%%%%%%%%%%%%%%%%%%%%%%%%%%%%%%%%%%%%%%%%%
\subsection{Bosonic truncation}
\label{subsec: truncation}

The bosonic model of \cite{Delduc:2017fib} was constructed making use of a gauge field.
Its form after eliminating the gauge field was also derived.
For the three-parameter deformation of $\mathbb{Z}_4$ permutation supercosets, we not only have a gauge field but also auxiliary fields.
This statement still holds for the bosonic truncation, where $\mApm^{(0)}$ plays the role of the gauge field and $\mApm^{(2)}$ is an auxiliary field.
Therefore, the simplest way to show that the bosonic truncation coincides with the model of \cite{Delduc:2017fib} is to compare the actions after eliminating the gauge and auxiliary fields.

An additional complication arises, however, as the action depending on the gauge field in \cite{Delduc:2017fib} is not written explicitly in terms of projection operators associated with the $\mathbb{Z}_2$ grading.
For this reason, in equation \eqref{eq:origLDHMT} of appendix \ref{apps: dhmt}, we first rewrite this action in the language used in this article.
We then eliminate the gauge field.
The action after eliminating the gauge field is given in \eqref{DHMT}.
The comparison between the bosonic truncation and \eqref{DHMT} will then be immediate and leads to the map between the parameters used in this article for the supercoset case and those used in \cite{Delduc:2017fib}.
It is worth noting that the bosonic action after eliminating the gauge field \eqref{DHMT} is written in a simpler form than in \cite{Delduc:2017fib}.

\paragraph{Action.}

When we consider the bosonic truncation the operator $d_-$ in \eqref{eq def dm}
becomes $d_- = 2 P_2$. Therefore, the operator $\mSm$ in \eqref{mSmresult2}
is now
\bea
\mSm &=& 2 \left( 1 - 2 P_2 k W \Omega_- \right) P_2 \left( 1 + 2 \Omega_- P_2 \right)^{-1} \nonumber \\
&=& 2 P_2 \left( 1 - 2 k W \Omega_- P_2 \right) \left( 1 + 2 \Omega_- P_2 \right)^{-1} .
\eea
The non-standard Wess-Zumino term \eqref{LWZ} becomes the standard
gauge invariant WZ term $S^B_{\wzk} [\gL^{\vphantom{-1}} \, \gR^{-1}]$, which is written in terms of one copy of the supergroup $\widehat{G}$.
Thus we have
\beq
\begin{split} \label{bos truncation}
S^B_{\eta_\smlr , \smk} [\gLR] =
&
\int d^2 \sigma \, \,
\STr \Bigl[
2 \, \mJp  P_2 \left( 1 - 2 k W \Omega_- P_2 \right)
\left( 1 + 2 \Omega_- P_2 \right)^{-1} \mJm
\Bigr] \\
& + S^B_{\wzk} [\gL^{\vphantom{-1}} \, \gR^{-1}] ,
\end{split}
\eeq
with $\Omega_-$ given in equation \eqref{Omgpm}.
We need to compare this action to the action \eqref{DHMT}, which is
reproduced here for convenience:
\beq
\begin{split}\label{DHMTtext}
\tilde{S}^B_{\tilde{\eta}_\smlr,\smtk}[\gLR] =
\tfrac12 \tN
\biggl ( &
\int d^2\sigma \, \, \STr
\Bigl[2 \, \mJp
P_2 \bigl( 1 - 2 \bt W \wOmgm P_2\bigr) \bigl( 1 + 2 \wOmgm P_2\bigr)^{-1} \mJm
\Bigr] \\
& + S^B_{\wztb} [ \gL^{\vphantom{-1}} \, \gR^{-1} ] \biggr ) ,
\end{split}
\eeq
where
\begin{align}\label{tOmgmtext}
\tN &= 2 \tk\bt^{-1}, \qquad \qquad
\wOmgm = \frac{1 + \tE^2}{2 \tilde{\alpha}^s_- (1 + \tE^2 + \tk W)^2}
\Bigl( - \tA \, \mRg - \frac{\tk W \tE^2}{1 + \tE^2} \mRg^2 \Bigr).
\end{align}
The values of $\bt$ and $\tilde{\alpha}^s_-$ in terms of the parameters $(\tk,\tELR)$ are given in equations \eqref{def b} and \eqref{defasaa}
respectively, while $\tA$ and $\tE$ are defined through \eqref{eq:mi2} and \eqref{eq:mi1}.

\paragraph{Map between the parameters.}

To determine the map between the parameters $(k,\ELR)$ and $(\tk,\tELR)$, we first
focus on the WZ terms in the actions \eqref{bos truncation} and \eqref{DHMTtext}. This
implies that we should identify $k$ and $\bt$, that is,
\beq
k = \bt=\frac{(2 + \tEL^2 + \tER^2) \tk}{(1 + \tEL^2) (1 + \tER^2) + \tk^2} .
\label{k to tk}
\eeq
It is then clear that we would have
\beq
\tilde{S}^B_{\tilde{\eta}_\smlr,\smtk}[g] =
\tfrac12 \tN
S^B_{{\eta}_\smlr,\smk}[g] ,
\eeq
if $ \wOmgm = \Omega_-$. Comparing the coefficients of
$\wOmgm$ in \eqref{tOmgmtext}
and of
$\Omega_-$ in \eqref{Omgpm}, we find the two conditions
\begin{align*}
\mu_{\L} - 1 &= \frac{\tEL^2 ( (1 + \tER^2)^2 - \tk^2)}{(2 + \tEL^2 + \tER^2)^2} , \\
\mu_{\R} - 1 &= \frac{\tER^2 ( (1 + \tEL^2)^2 - \tk^2)}{(2 + \tEL^2 + \tER^2)^2} .
\end{align*}
Using the form of $\mu$ given in equation \eqref{ansatz}, these relations
may be rewritten as
\bse \label{id-param}
\bea
\frac{(1 - k^2) \EL^2}{(1 - \EL^2)(1 - \ER^2) - k^2 (1 - \EL^2 - \ER^2)} &=& \frac{\tEL^2 ( (1 + \tER^2)^2 - \tk^2)}{(2 + \tEL^2 + \tER^2)^2} , \\
\frac{(1 - k^2) \ER^2}{(1 - \EL^2)(1 - \ER^2) - k^2 (1 - \EL^2 - \ER^2)} &=& \frac{\tER^2 ( (1 + \tEL^2)^2 - \tk^2)}{(2 + \tEL^2 + \tER^2)^2} .
\eea
\ese
Therefore, the map
between the parameters used in the present article and those
used in \cite{Delduc:2017fib} is defined by \eqref{k to tk} and \eqref{id-param}. This demonstrates
the agreement between the bosonic truncation of the three-parameter deformation of $\mathbb{Z}_4$ permutation supercosets
and the model constructed in \cite{Delduc:2017fib}.

%%%%%%%%%%%%%%%%%%%%%%%%%%%%%%%%%%%%%%%%%%%%%%%%%%%%%%%%%%%%%%%%%%%%%%%%%%%%%%%%
\subsection{Mixed flux model}
\label{subsec: mf}

Now let us investigate the limit in which we expect to recover the mixed flux model of \cite{Cagnazzo:2012se}.
For this we take $\ELR = 0$.
It then immediately follows from \eqref{ansatz} and \eqref{Omgpm} that $\mu= 1$ and $\Omega_-=0$.
As a consequence, the relation \eqref{mSmresult2} becomes $\mSm = 2 \, \curlP$.
We also have from equation \eqref{ansatz lambda} that
$\lambda=\sqrt{1 - k^2}$. It remains to compute the values
of $d_-$ and $\curlP$, defined in \eqref{eq def dm} and
\eqref{eq: defcurlP} respectively,
when $\ELR = 0$. Doing so we find
\beq \label{eq:projlimitmixedflux}
\curlP(\ELR = 0,k) = \tfrac12 \bigl( d_-
+ P_F \, k W \left( 1 - d_- \right) \bigr)= P_2 + \tfrac12 \sqrt{1-k^2} ( P_1 - P_3 ).
\eeq
Therefore, we have
\beqz
S_{\eta_\smlr=0 , \smk}[g] =
2\biggl(
\int d^2 \sigma \, \,
\STr \Bigl[ \mJp \, \curlP \mJm \Bigr] + \tfrac12 S_{\mfk} [g]
\biggr).
\eeqz
This action indeed corresponds to the mixed flux model of \cite{Cagnazzo:2012se}, written
in the form given in \cite{Hoare:2013pma,Babichenko:2014yaa,Hoare:2014oua}.

%%%%%%%%%%%%%%%%%%%%%%%%%%%%%%%%%%%%%%%%%%%%%%%%%%%%%%%%%%%%%%%%%%%%%%%%%%%%%%%%
\subsection{Bi-Yang-Baxter deformation}
\label{subsec: sbyb}
The limit that should correspond to the bi-Yang-Baxter deformation of $\mathbb{Z}_4$ permutation supercosets is given by taking $k = 0$.
We then have
\begin{align}
\lambda &= \sqrt{( 1 -\EL^2 ) ( 1 -\ER^2 )}, \qquad \qquad \muLR = 1+ \frac{\ELR^2}{\lambda^2}, \label{eq:valmubyb}
\end{align}
and
\begin{align}
d_- &= 2 P_2 + \lambda (P_1 - P_3), \qquad \qquad
\curlP = \tfrac12 d_-, \qquad \qquad
\mSm = 2 \curlP (1+ 2 \Omega_- \curlP)^{-1}. \label{eqvalombyb}
\end{align}
Therefore,
\beq \label{curlPbyblim}
\curlP(\ELR,k=0) = P_2 + \frac{\sqrt{(1-\EL^2)(1-\ER^2)}}{2} ( P_1 - P_3 ).
\eeq
It remains to compute $\Omega_-$ when $k=0$. Starting from
\eqref{Omgpm}, we find
\beqz
\Omega_-(\ELR,k=0)=- \sqrt{\mu - 1} \, \mRg = - \tfrac12 \vk \, \mRg,
\eeqz
where we have used equation \eqref{eq:valmubyb} and defined
\beqz
\vk = \frac{2}{\sqrt{(1-\EL^2)(1-\ER^2)}} \textrm{diag} (\EL,\ER).
\eeqz
Substituting this expression for $\Omega_-$ into $\mSm$ \eqref{eqvalombyb} we find that the action \eqref{Lfullresult2} becomes
\beqz
S_{\eta_\smlr , \smk=0 }[g] =
2
\int d^2 \sigma \, \,
\STr\Bigl( \mJp \, \curlP \left(1 - \vk \, \mRg \, \curlP \right)^{-1} \mJm \Bigr) ,
\eeqz
with $\curlP$ given in equation \eqref{curlPbyblim}.
This action indeed coincides with the bi-Yang-Baxter deformation of $\mathbb{Z}_4$ permutation
supercosets constructed in \cite{Hoare:2014oua}.

%%%%%%%%%%%%%%%%%%%%%%%%%%%%%%%%%%%%%%%%%%%%%%%%%%%%%%%%%%%%%%%%%%%%%%%%%%%%%%%%
\section{Conclusion}
\label{sec: conclusion}

In this article we have constructed the bi-Yang-Baxter deformation of the mixed flux model of \cite{Cagnazzo:2012se} giving a three-parameter integrable deformation of the semi-symmetric space $\sigma$-model on $\mathbb{Z}_4$ permutation supercosets.
Furthermore, we demonstrated its classical integrability via the existence of a Lax pair and confirmed the agreement of various truncations and limits with known models.

For $\widehat{F} = PSU(1,1|2)$ or $\widehat{F} = D(2,1;\alpha)$ the mixed flux model, together with the appropriate number of free compact bosons, is a $\kappa$-symmetry gauge-fixing of the Green-Schwarz superstring on AdS$_3 \times$ S$^3 \times$ T$^4$ or AdS$_3 \times$ S$^3 \times$ S$^3 \times$ S$^1$ supported by mixed R-R and NS-NS flux \cite{Babichenko:2009dk,Cagnazzo:2012se,Wulff:2014kja,Wulff:2015mwa}.
Yang-Baxter deformations based on solutions of the modified classical Yang-Baxter equation do not typically describe strings on type II supergravity backgrounds \cite{Arutyunov:2015qva,Borsato:2016ose}.
Instead the background fields satisfy a generalisation of the supergravity equations \cite{Arutyunov:2015mqj,Wulff:2016tju}.
This is indeed the case for the Yang-Baxter and bi-Yang-Baxter deformations of the AdS$_3 \times$ S$^3 \times$ T$^4$ background supported by pure R-R flux \cite{Hoare:2015wia,Arutyunov:2015mqj,Araujo:2018rbc}.
It would be interesting to determine the R-R fluxes that support the metrics and B-fields of the three-parameter deformations of AdS$_3 \times$ S$^3 \times$ T$^4$ and AdS$_3 \times$ S$^3 \times$ S$^3 \times$ S$^1$ (see \cite{Lukyanov:2012zt,Delduc:2017fib} and appendix \ref{app:mbads3s3}) and confirm that the generalised supergravity equations are satisfied.
Furthermore, the generalised supergravity equations should imply that the model is scale invariant and UV finite on a flat two-dimensional worldsheet \cite{Arutyunov:2015mqj}.
It would be important to confirm that this is indeed the case, for example, by checking the vanishing of the one-loop beta function for the $\sigma$-model coupling.

While Yang-Baxter deformations based on solutions of the modified classical Yang-Baxter equation do not typically describe strings on supergravity backgrounds, their T-duals \cite{Hoare:2015wia,Arutyunov:2015mqj} and Poisson-Lie duals \cite{Hoare:2018ebg} can.
Poisson-Lie duality, introduced in \cite{Klimcik:1995ux,Klimcik:1995jn,Klimcik:1995dy}, is a generalisation of (non-abelian) T-duality to models with Poisson-Lie symmetry.
Poisson-Lie duals of the Yang-Baxter deformation of the principal chiral model plus Wess-Zumino term have been studied in \cite{Klimcik:2017ken,Demulder:2017zhz}.
Extending this analysis to the bi-Yang-Baxter case, as well as to the R-R sector (see for instance \cite{Hoare:2018ebg,Severa:2018pag,Demulder:2018lmj}) would be necessary for investigating whether there exist duals of the three-parameter deformations of the AdS$_3 \times$ S$^3 \times$ T$^4$ and AdS$_3 \times$ S$^3 \times$ S$^3 \times$ S$^1$ backgrounds that are solutions of type II supergravity.

In order to understand the possible Poisson-Lie duals of the three-parameter integrable deformation it would also be helpful to study the Poisson-Lie symmetry of the model, together with the associated $q$-deformation of the global symmetry algebra \cite{Delduc:2016ihq}.
An alternative route to exploring the $q$-deformed symmetry would be to compute the light-cone gauge dispersion relation and S-matrix as done for the Yang-Baxter deformation of the AdS$_5 \times$ S$^5$ superstring in \cite{Arutyunov:2013ega,Arutyunov:2015qva}.
An initial proposal, based on symmetry considerations, for the deformed dispersion relation and S-matrix (up to overall phases) in the massive sector was given in \cite{Hoare:2014oua,Regelskis:2015xxa} following \cite{Beisert:2008tw}.
These are deformations of the undeformed dispersion relation and S-matrix constructed in \cite{Borsato:2012ud,Borsato:2013qpa,Borsato:2014hja}.

Finally, it would also be interesting to perform the Hamiltonian analysis of 
this integrable $\sigma$-model and determine its twist function 
\cite{Vicedo:2010qd} (see \cite{Lacroix:2018njs} for a review).
This would be the first step towards its reinterpretation as an affine Gaudin model, in the spirit of \cite{Vicedo:2017cge}.

\paragraph{Acknowledgments.}

This work is partially supported by the French Agence Nationale de la Recherche (ANR) under grant ANR-15-CE31-0006 DefIS.
BH is partially supported by grant no. 615203 from the European Research Council under the FP7.

%%%%%%%%%%%%%%%%%%%%%%%%%%%%%%%%%%%%%%%%%%%%%%%%%%%%%%%%%%%%%%%%%%%%%%%%%%%%%%%%
\appendix

%%%%%%%%%%%%%%%%%%%%%%%%%%%%%%%%%%%%%%%%%%%%%%%%%%%%%%%%%%%%%%%%%%%%%%%%%%%%%%%%
\section{Coefficients of the linear maps used for the Lax pair}
\label{app: coeffs}
\def\sigmamark{}

In this appendix we give the values of the various coefficients
used in equations \eqref{eq:linmapA} and \eqref{eq:eezz}. We have
\begingroup
\allowdisplaybreaks
\begin{align*}
& b_\pm^0 = 1 , \qquad b_\pm^2 = \frac{ - (1-\etaL^2)(1-\etaR^2) + k^2(1-\etaL^2-\etaR^2)}{\sqrt{1-k^2}(1-k^2\etaL^2\etaR^2)} ,
\\
& c_\pm^0 = 0 , \qquad c_\pm^2 = \frac{ - (1-k^2)(\etaL^2 - \etaR^2) \pm k (1-k^2 + \etaL^2\etaR^2)}{\sqrt{1-k^2}(1-k^2 -\etaL^2\etaR^2)} ,
\\
& \begin{aligned}
& b_+^1 = \mathbf{a}(k,\etaL,\etaR) + \mathbf{a}(-k,\etaR,\etaL) , \qquad &
& b_-^1 = \mathbf{d}(k,\etaL,\etaR) + \mathbf{d}(-k,\etaR,\etaL) ,
\\
& b_+^3 = \mathbf{d}(-k,\etaL,\etaR) + \mathbf{d}(k,\etaR,\etaL) , \qquad &
& b_-^3 = \mathbf{a}(-k,\etaL,\etaR) + \mathbf{a}(k,\etaR,\etaL) ,
\\
& c_+^1 = \mathbf{a}(k,\etaL,\etaR) - \mathbf{a}(-k,\etaR,\etaL) , \qquad &
& c_-^1 = - \mathbf{d}(k,\etaL,\etaR) + \mathbf{d}(-k,\etaR,\etaL) ,
\\
& c_+^3 = - \mathbf{d}(-k,\etaL,\etaR) + \mathbf{d}(k,\etaR,\etaL) , \qquad &
& c_-^3 = \mathbf{a}(-k,\etaL,\etaR) - \mathbf{a}(k,\etaR,\etaL) ,
\end{aligned}
\\
& \mathbf{a}(k,\etaL,\etaR) = - \left(\frac{1-k}{1+k}\right)^{\frac{3}{4}} \frac{(1-k^2 -\etaL^2)(\etaR^2 - 2(1+k)) + k(1-k^2)}{2\sqrt{1-k^2}\sqrt{1-k^2 -\etaL^2\etaR^2}\sqrt{1-k^2-\etaL^2}} ,
\\
& \mathbf{d}(k,\etaL,\etaR) = \left(\frac{1-k}{1+k}\right)^{\frac{1}{4}}
\frac{(1-k^2 -\etaR^2)\etaL^2 + k(1-k^2)}{2\sqrt{1-k^2}\sqrt{1-k^2 -\etaL^2\etaR^2}\sqrt{1-k^2-\etaR^2}} ,
\end{align*}%
\endgroup
while for the remaining coefficients we find
\begingroup
\allowdisplaybreaks
\small
\begin{align*}
&
\begin{aligned}
& e_2 = \frac{\sqrt{1-k^2}(1-k^2-\etaL^2\etaR^2)}{2\big((1-\etaL^2)(1-\etaR^2) - k^2 (1-\etaL^2-\etaR^2)\big)} , \qquad
&& \tilde{f}_2 = \frac{k(1-k^2 +\etaL^2 \etaR^2)}{2\big((1-\etaL^2)(1-\etaR^2) - k^2 (1-\etaL^2-\etaR^2)\big)} ,
\\
& z_0 = \frac{(1-k^2)\big((1-\etaL^2)(1-\etaR^2) - k^2\big)}{(1-\etaL^2)(1-\etaR^2) - k^2 (1-\etaL^2-\etaR^2)} ,
&& \tilde{w}_0 = \frac{-k\sqrt{1-k^2}(1-k^2-\etaL^2\etaR^2)}{(1-\etaL^2)(1-\etaR^2) - k^2 (1-\etaL^2-\etaR^2)} ,
\\
& z_2 = \frac{\sqrt{1-k^2}(1-k^2-\etaL^2\etaR^2)}{(1-\etaL^2)(1-\etaR^2) - k^2 (1-\etaL^2-\etaR^2)} ,
&& \tilde{z}_2 = \frac{(1-k^2)(\etaL^2 - \etaR^2)}{(1-\etaL^2)(1-\etaR^2) - k^2 (1-\etaL^2-\etaR^2)} ,
\\
& f_2 = w_0 = w_2 = 0,
&& \tilde{e}_2 = \tilde{z}_0 = \tilde{w}_2 = 0,
\\
& e_1 = \mathbf{e}(k,\etaL,\etaR) + \mathbf{e}(-k,\etaR,\etaL) ,
&& \tilde{e}_1 = \mathbf{e}(-k,\etaL,\etaR) - \mathbf{e}(k,\etaR,\etaL) ,
\\
& e_3 = - \mathbf{e}(-k,\etaL,\etaR) - \mathbf{e}(k,\etaR,\etaL) ,
&& \tilde{e}_3 = - \mathbf{e}(k,\etaL,\etaR) + \mathbf{e}(-k,\etaR,\etaL) ,
\\
& w_1 = \mathbf{w}(k,\etaL,\etaR) + \mathbf{w}(-k,\etaR,\etaL) ,
&& \tilde{w}_1 = - \mathbf{w}(-k,\etaL,\etaR) + \mathbf{w}(k,\etaR,\etaL) ,
\\
& w_3 = \mathbf{w}(-k,\etaL,\etaR) + \mathbf{w}(k,\etaR,\etaL) ,
&& \tilde{w}_3 = - \mathbf{w}(k,\etaL,\etaR) + \mathbf{w}(-k,\etaR,\etaL) ,
\\
& z_1 = \mathbf{z}(k,\etaL,\etaR) + \mathbf{z}(-k,\etaR,\etaL) ,
&& \tilde{z}_1 = \mathbf{z}(-k,\etaL,\etaR) - \mathbf{z}(k,\etaR,\etaL) ,
\\
& z_3 = \mathbf{z}(-k,\etaL,\etaR) + \mathbf{z}(k,\etaR,\etaL) ,
&& \tilde{z}_3 = \mathbf{z}(k,\etaL,\etaR) - \mathbf{z}(-k,\etaR,\etaL) ,
\\
& f_1 = \mathbf{f}(k,\etaL,\etaR) + \mathbf{f}(-k,\etaR,\etaL) ,
&& \tilde{f}_1 = - \mathbf{f}(-k,\etaL,\etaR) + \mathbf{f}(k,\etaR,\etaL) ,
\\
& f_3 = - \mathbf{f}(-k,\etaL,\etaR) - \mathbf{f}(k,\etaR,\etaL) ,
&& \tilde{f}_3 = \mathbf{f}(k,\etaL,\etaR) - \mathbf{f}(-k,\etaR,\etaL) ,
\end{aligned}
\\
& \mathbf{e}(k,\etaL,\etaR) = -\left(\frac{1-k}{1+k}\right)^{\frac14}\frac{(1-k^2)(1-k^2-\etaL^2\etaR^2)^{\frac{3}{2}}\big(1-k^2-(1+k)\etaR^2\big)}{8\sqrt{1-k^2-\etaR^2}\big((1-\etaL^2)(1-\etaR^2)-k^2(1-\etaL^2-\etaR^2)\big)^2 } ,
\\
& \mathbf{w}(k,\etaL,\etaR) = \left(\frac{1-k}{1+k}\right)^{\frac14}\frac{(1-k^2)(1-k^2-\etaL^2\etaR^2)^{\frac{3}{2}}}{8\sqrt{1-k^2-\etaR^2}\big((1-\etaL^2)(1-\etaR^2)-k^2(1-\etaL^2-\etaR^2)\big)} ,
\\
& \mathbf{z}(k,\etaL,\etaR) = \left(\frac{1-k}{1+k}\right)^{\frac14}\frac{\sqrt{1-k^2}\sqrt{1-k^2-\etaL^2\etaR^2}\big(2(1+k)(1-k^2-\etaR^2) - k (1-k^2-\etaL^2\etaR^2)\big)}{4\sqrt{1-k^2-\etaL^2}\big((1-\etaL^2)(1-\etaR^2)-k^2(1-\etaL^2-\etaR^2)\big)} ,
\\
& \mathbf{f}(k,\etaL,\etaR) = \left(\frac{1-k}{1+k}\right)^{\frac14} \frac{\sqrt{1-k^2}\sqrt{1-k^2-\etaL^2\etaR^2}}
{4\sqrt{1-k^2-\etaL^2}\big((1-\etaL^2)(1-\etaR^2)-k^2(1-\etaL^2-\etaR^2)\big)^2} \times
\\ &
\big(k(1-k^2)^2 - k(1-k^2)(3+k)\etaL^2 - 2(1-k^2)(1+k) \etaR^2 +(1+k)(4+k-k^2)\etaL^2\etaR^2-(2+k(1+k))\etaL^4\etaR^2\big) .
\end{align*}%
\endgroup

%%%%%%%%%%%%%%%%%%%%%%%%%%%%%%%%%%%%%%%%%%%%%%%%%%%%%%%%%%%%%%%%%%%%%%%%%%%%%%%%
\section{Three-parameter deformation of \texorpdfstring{$\mathbb{Z}_2$}{Z2} permutation cosets}
\label{apps: dhmt}

In this appendix we rewrite the three-parameter deformation of $\mathbb{Z}_2$ permutation cosets constructed in \cite{Delduc:2017fib} in the language used in this article.
We first give the action which includes a gauge field and then eliminate this gauge field.
Note that, to be precise, we write the action of the $\mathbb{Z}_2$ permutation coset embedded in a $\mathbb{Z}_4$ permutation supercoset.
This explains why the action is written using the supertrace as opposed to the negative trace, which would be the appropriate bilinear form for compact Lie groups.

\paragraph{Action with gauge field.}

The three parameters are denoted by $\tk$ and $\tELR$.
The action is
\begin{align}
\tilde{S}^B_{\tilde{\eta}_\smlr,\smtk} [g, \wmApm^{(0)}]
= &\int d^2\sigma \, \, \STr \Bigl[
\bigl( \mJp - \wmAp^{(0)} \bigr) \wmOm \bigl( \mJm - \wmAm^{(0)} \bigr)
\nonumber \\
& \qquad \qquad \quad + \bigl( \mJp - \wmAp^{(0)} \bigr) \tk W \mJm^{(2)} - \tk W \mJp^{(2)}
\bigl( \mJm - \wmAm^{(0)} \bigr)
\Bigr]
\label{eq:origLDHMT}
\\
&+ S^B_{\wztk}[\gL^{\vphantom{-1}} \, \gR^{-1} ] . \nonumber
\end{align}

The operator $\wmOm$ is defined as $\wmOm = \text{diag} (\wOmL , \wOmR)$ with
\beqz
\wOmL = (1 + \tEL^2) \, \frac{1 + \tA_{\L} R_{\gL}}{1 - \tEL^2 R_{\gL}^2} ,
\qquad
\wOmR = (1 + \tER^2) \, \frac{1 + \tA_{\R} R_{\gR}}{1 - \tER^2 R_{\gR}^2}.
\eeqz
The coefficients $\tilde{A}_{\L, \R}$ take the values \cite{Kawaguchi:2011mz,
Kawaguchi:2013gma,Delduc:2014uaa,Delduc:2017fib}
\beq\label{eq:mi1}
\tA_{\L} = \tEL \sqrt{ 1 -\frac{\tk^2}{1 + \tEL^2} } , \qquad
\tA_{\R} = \tER \sqrt{ 1 -\frac{\tk^2}{1 + \tER^2} } .
\eeq

\paragraph{Action after eliminating the gauge field.}

The equation of motion for the gauge field $\wmAmp^{(0)}$ is
\beq
P_0 \, \wmOpm \bigl( \mJpm - \wmApm^{(0)} \bigl) \mp \tk W \mJpm^{(2)} = 0 ,
\label{eom A DHMT}
\eeq
with $\wmOp = (\wmOm)^t$. Let us introduce the current $\wmQpm$ defined as
\beq \label{def qtapp}
\wmQpm = \bigl( \wmOpm \mp \tk W \bigr) \bigl( \mJpm - \wmApm^{(0)} \bigr).
\eeq
The equation of motion \eqref{eom A DHMT} implies
that $P_0 \, \wmQpm = 0$ and thus $P_2 \, \wmQpm = \wmQpm$.
Therefore,
\bse \label{A DHMT}
\bea
\mJpm^{(0)} - \wmApm^{(0)} &=& P_0 \bigl( \wmOpm \mp \tk W \bigr)^{-1} \wmQpm^{(2)} , \\
\mJpm^{(2)} &=& P_2 \bigl( \wmOpm \mp \tk W \bigr)^{-1} \wmQpm^{(2)} .
\eea
\ese
It remains to compute the inverse operators appearing in these
equations. This may be done by using the formula \eqref{eq:formulainversionR}.
Doing so we find
\beq
\bigl( \wmOpm \mp \tk W \bigr)^{-1} = \frac{1}{1 + \tE^2 \mp \tk W} \, \Pi_g +
\frac{1 + \tE^2}{(1 + \tE^2 \mp \tk W)^2} \left( \pm \tA \, \mRg - ( 1 \mp \tk W ) \, \mRg^2 \right) ,
\label{O mp kW DHMT}
\eeq
where
\beq\label{eq:mi2}
\tE = \text{diag} ( \tEL , \tER ) \quad \text{and} \quad \tA = \text{diag} (\tA_{\L} , \tA_{\R}).
\eeq
We then write $\Pi_g = 1 + \mRg^2$ and decompose the coefficient multiplying $\Pi_g$
into its symmetric and anti-symmetric part, that is
\beqz
\frac{1}{1 + \tE^2 \mp \tk W} = \tilde{\alpha}^s_\pm 1+ \tilde{\alpha}^a_\pm W ,
\eeqz
with
\beq \label{defasaa}
\tilde{\alpha}^s_{\pm} = \frac{2 + \tEL^2 + \tER^2}{2 ( 1 + \tEL^2 \mp \tk ) ( 1 + \tER^2 \pm \tk )} , \qquad
\tilde{\alpha}^a_{\pm} = - \frac{\tEL^2 - \tER^2 \mp 2\tk}{2 ( 1 + \tEL^2 \mp \tk ) ( 1 + \tER^2 \pm \tk )} .
\eeq
Finally we find
\beq
\bigl( \wmOpm \mp \tk W \bigr)^{-1}
= \tilde{\alpha}^s_{\pm} \Bigl( 1 + \frac{\tilde{\alpha}^a_{\pm}}{\tilde{\alpha}^s_{\pm}} \, W + 2 \wOmgpm \Bigr) ,
\label{O mp kW DHMT 2}
\eeq
where the operators $\wOmgpm$ are given by
\beq
2 \tilde{\alpha}^s_{\pm} \wOmgpm = \frac{1 + \tE^2}{(1 + \tE^2 \mp \tk W)^2} \Bigl( \pm \tA \, \mRg \pm \frac{\tk W \tE^2}{1 + \tE^2} \, \mRg^2 \Bigr) .
\eeq
Equations \eqref{A DHMT} may then be expressed as
\bse
\bea
\mJpm^{(0)} - \wmApm^{(0)} &=& \tilde{\alpha}^s_\pm \Bigl( \frac{\tilde{\alpha}^a_\pm}{\tilde{\alpha}^s_\pm} \, W
+ 2 P_0 \wOmgpm \Bigr) \wmQpm^{(2)} , \label{appaoneeq} \\
\mJpm^{(2)} &=&\tilde{\alpha}^s_{\pm} \bigl( 1 + 2 P_2 \wOmgpm \bigr) \wmQpm^{(2)} . \label{eq:relatJandQt}
\eea
\ese

We can now compute the action obtained after eliminating the gauge field
$\wmApm^{(0)}$. We first eliminate $\wmAm^{(0)}$ by making use
of the equation of motion of $\wmAp^{(0)}$. The Lagrangian defined by the first two lines
of the action \eqref{eq:origLDHMT} becomes $\STr(\mJp \tilde{{\mathcal C}}_-)$ with
\beq
\tilde{{\mathcal C}}_-= P_2 \Bigl( \bigl(\wmOm + \tk W \bigr) \bigl( \mJm -
\wmAm^{(0)} \bigr) - 2 \tk W \bigl( \mJm - \wmAm^{(0)} \bigr) \Bigr).
\eeq
The quantity $\tilde{{\mathcal C}}_-$ can be written as
\begin{equation*}
\tilde{{\mathcal C}}_-=\wmQm^{(2)} - 2 \tilde{\alpha}^s_- \tk W \Bigl( \frac{\tilde{\alpha}^a_-}
{\tilde{\alpha}^s_-} \, W +
2 P_0 \wOmgm \Bigr) \wmQm^{(2)} \nonumber \\
= \frac{(1 + \tEL^2)(1 + \tER^2) + \tk^2}{(1 + \tEL^2 + \tk) (1 + \tER^2 - \tk)} \Bigl(
1 -2 \bt W P_0 \wOmgm \Bigr) \wmQm^{(2)} ,
\end{equation*}
where we have made use of the definition \eqref{def qtapp}, the property
$P_2 W \mJm = P_2 W \mJm^{(0)}$
and equations \eqref{appaoneeq}. We have also defined
\beq
\bt = \frac{(2 + \tEL^2 + \tER^2) \, \tk}{(1 + \tEL^2)(1 + \tER^2) + \tk^2} .
\label{def b}
\eeq
The last step is to invert the relation \eqref{eq:relatJandQt} and use the identities
\bse
\bea
\bigl( 1 + 2 P_2 \wOmgm \bigr)^{-1} \, P_2 &=& P_2 \bigl( 1 + 2 \wOmgm P_2\bigr)^{-1} , \\
\bigl( 1 - 2 \bt W P_0 \wOmgm \bigr) P_2 &=& P_2 \bigl( 1 - 2 \bt W \wOmgm P_2\bigr) ,
\eea
\ese
to write the action after elimination of the gauge field as
\beq
\begin{split}\label{DHMT}
\tilde{S}^B_{\tilde{\eta}_\smlr,\smtk}[\gLR] =
\tfrac12 \tN
\biggl ( &
\int d^2\sigma \, \, \STr
\Bigl[2 \, \mJp
P_2 \bigl( 1 - 2 \bt W \wOmgm P_2\bigr) \bigl( 1 + 2 \wOmgm P_2\bigr)^{-1} \mJm
\Bigr] \\
& + S^B_{\wztb} [ \gL^{\vphantom{-1}} \, \gR^{-1} ] \biggr ) ,
\end{split}
\eeq
where
\begin{align}\label{tOmgm}
\tN &= 2 \tk\bt^{-1}, \qquad \qquad
\wOmgm = \frac{1 + \tE^2}{2 \tilde{\alpha}^s_- (1 + \tE^2 + \tk W)^2}
\Bigl( - \tA \, \mRg - \frac{\tk W \tE^2}{1 + \tE^2} \mRg^2 \Bigr).
\end{align}

%%%%%%%%%%%%%%%%%%%%%%%%%%%%%%%%%%%%%%%%%%%%%%%%%%%%%%%%%%%%%%%%%%%%%%%%%%%%%%%%
\section{Metric and B-field for the three-parameter deformation of \texorpdfstring{S$^3$}{S3}}\label{app:mbads3s3}

In this appendix we give simple expressions for the metric and B-field in the case of
the three-parameter deformation of S$^3$. Other expressions have been previously
obtained in \cite{Lukyanov:2012zt,Delduc:2017fib}. We start from the result \eqref{DHMT}
with the supertrace being replaced by the negative trace.
In the case of $G = SU(2)$, we take the familiar basis for the Lie algebra $\mathfrak{su}(2)$
\beqz
T_a = \frac{i}{2}\sigma_a ,
\eeqz
where $\sigma_a$ are the Pauli matrices.
The Drinfel'd-Jimbo \cite{Drinfeld:1985rx,Jimbo:1985zk,Belavin:1984} solution of the modified classical Yang-Baxter equation is then given by
\beqz
R T_1 = - T_2 , \qquad R T_2 = T_1 , \qquad R T_3 = 0.
\eeqz
We use the gauge symmetry $\gLR \to \gLR g_{0}$ to fix $\gR = 1$ and parametrise $\gL$ as
\beqz
\gL = \exp[(-\varphi+\phi)T_3)].(r 1 - 2 \sqrt{1-r^2} T_1).\exp[(\varphi+\phi)T_3)] .
\eeqz

We substitute into the action \eqref{bos truncation} and replace $\STr$ by $-\Tr$ to arrive at the three-parameter
($\tEL$, $\tER$ and $\tk$) deformation of the $\mathrm{S}^3$ $\sigma$-model.
Comparing with the usual $\sigma$-model form in conformal gauge
\beqz
- 2 \int d^2 \sigma \, \, (\eta^{\alpha\beta} G_{\mu\nu} \partial_\alpha X^\mu
\partial_\beta X^\nu + \epsilon^{\alpha\beta} B_{\mu\nu} \partial_\alpha X^\mu \partial_\beta X^\nu)
= 2 \int d^2 \sigma \, \, (G_{\mu\nu} + B_{\mu\nu}) \partial_+ X^\mu \partial_- X^\nu,
\eeqz
we find the following target space metric and B-field
\beqz\begin{split}
ds^2 & = \tilde{F}^{-1} \Big( \frac{(1+\tilde{q}^2r^2(1-r^2))}{1-r^2} dr^2 -2
\tilde{q} \tilde{\kappa}_- (1-r^2) r drd\varphi - 2 \tilde{q} \tilde{\kappa}_+ r^3 dr d\phi
\\ & \qquad \qquad + (1+\tilde{\kappa}_-^2(1-r^2))(1-r^2)d\varphi^2 + (1+\tilde{\kappa}_+^2 r^2)r^2
d\phi^2 + 2\tilde{\kappa}_+\tilde{\kappa}_- r^2 (1-r^2) d\varphi d\phi \Big) ,
\\
B & = \tilde{a} \tilde{F}^{-1} \big(\tilde{\kappa}_+^2 - \tilde{\kappa}_-^2 + \tilde{q}^2(1-2r^2)\big)
(- \tilde{\kappa}_+ rdr \wedge d\varphi + \tilde{\kappa}_- rdr \wedge d\phi) 
\\ & \qquad \qquad - \tilde{a} \tilde{q} \tilde{F}^{-1} \big(1-2r^2 - \tilde{\kappa}_+^2 r^4 +
\tilde{\kappa}_-^2 (1-r^2)^2 \big) d\varphi \wedge d\phi,
\\
\tilde{F} & = 1+ \tilde{\kappa}_+^2 r^2 + \tilde{\kappa}_-^2 (1-r^2) + \tilde{q}^2 r^2(1-r^2) , \qquad
\tilde{a} = \frac{1}{\sqrt{(\tilde{q}^2 + \tilde{\kappa}_+^2 +
\tilde{\kappa}_-^2)^2 + 4 (\tilde{q}^2 - \tilde{\kappa}_+^2 \tilde{\kappa}_-^2)}} ,
\end{split}\eeqz
where we have introduced the new parameters
\beqz
\tilde{\kappa}_\pm = \sqrt{\frac{(4+(\tA_{\L} \pm \tA_{\R})^2)(1+\tEL^2)
(1+\tER^2)}{(2+\tEL^2+\tER^2)^2} - 1} ,
\qquad \tilde{q} = \frac{2 \tEL \tER \tk}{2+\tEL^2+\tER^2}.
\eeqz
For convenience when relating the two sets of parameters we
assume that $\tEL$, $\tER$ and $\tk$ are all positive and all square roots are positive so that
$\tilde{\kappa}_\pm$, $\tilde{q}$ and $\tilde{a}$ are also positive.
Furthermore, we assume that we are in a region of parameter space such that the following relations hold
\beqz\begin{split}
\tilde{\kappa}_+
\tilde{\kappa}_- & = \frac{(\tER^2 - \tEL^2)
(1-\tEL^2\tER^2 - \tk^2)}{(2+\tEL^2+\tER^2)^2} ,
\\
\tilde{q} \tilde{\kappa}_\pm & =
\frac{2 \tk (\tA_{\L}(1+\tEL^2)\tER^2 \pm \tA_{\R} (1+\tER^2)\tEL^2)}{(2+\tEL^2+\tER^2)^2} ,
\\
\tilde{a} & = \frac{(2+\tEL^2+\tER^2)^2}{4\tEL \tER((1+\tEL^2)(1+\tER^2)+\tk^2)} .
\end{split}\eeqz

Let us conclude with some observations.
First, we note that the first line of the B-field is closed and hence locally it can be set to zero by a gauge transformation.
Second, the metric is invariant under the following formal transformation
\beqz
r \to \sqrt{1-r^2} , \quad \varphi \leftrightarrow \phi , \qquad \tilde{\kappa}_+ \leftrightarrow
\tilde{\kappa}_- , \qquad \tilde{q} \to -\tilde{q} ,
\eeqz
while the B-field changes just by an overall sign.
Finally, the analogous deformation of AdS$_3$ can be found by analytically continuing
\begin{equation*}
\rho \to i r, \qquad \varphi \to t, \qquad \phi \to \psi,
\end{equation*}
as well as flipping the overall sign of the metric and B-field.

\providecommand{\href}[2]{#2}\begingroup\raggedright\endgroup


\begin{thebibliography}{10}

\bibitem{Delduc:2017fib}
F.~Delduc, B.~Hoare, T.~Kameyama and M.~Magro, \emph{{Combining the
  bi-Yang-Baxter deformation, the Wess-Zumino term and TsT transformations in
  one integrable $\sigma$-model}},
  \href{https://doi.org/10.1007/JHEP10(2017)212}{\emph{JHEP} {\bfseries 10}
  (2017) 212} [\href{https://arxiv.org/abs/1707.08371}{{\ttfamily
  1707.08371}}].

\bibitem{Lukyanov:2012zt}
S.~L. Lukyanov, \emph{{The integrable harmonic map problem versus Ricci flow}},
  \href{https://doi.org/10.1016/j.nuclphysb.2012.08.002}{\emph{Nucl. Phys.}
  {\bfseries B865} (2012) 308}
  [\href{https://arxiv.org/abs/1205.3201}{{\ttfamily 1205.3201}}].

\bibitem{Klimcik:2002zj}
C.~Klim\v{c}\'{i}k, \emph{{Yang-Baxter sigma models and dS/AdS T duality}},
  \href{https://doi.org/10.1088/1126-6708/2002/12/051}{\emph{JHEP} {\bfseries
  12} (2002) 051} [\href{https://arxiv.org/abs/hep-th/0210095}{{\ttfamily
  hep-th/0210095}}].

\bibitem{Cherednik:1981df}
I.~V. Cherednik, \emph{{Relativistically Invariant Quasiclassical Limits of
  Integrable Two-dimensional Quantum Models}},
  \href{https://doi.org/10.1007/BF01086395}{\emph{Theor. Math. Phys.}
  {\bfseries 47} (1981) 422}.

\bibitem{Klimcik:2008eq}
C.~Klim\v{c}\'{i}k, \emph{{On integrability of the Yang-Baxter sigma-model}},
  \href{https://doi.org/10.1063/1.3116242}{\emph{J. Math. Phys.} {\bfseries 50}
  (2009) 043508} [\href{https://arxiv.org/abs/0802.3518}{{\ttfamily
  0802.3518}}].

\bibitem{Delduc:2013fga}
F.~Delduc, M.~Magro and B.~Vicedo, \emph{{On classical $q$-deformations of
  integrable sigma-models}},
  \href{https://doi.org/10.1007/JHEP11(2013)192}{\emph{JHEP} {\bfseries 11}
  (2013) 192} [\href{https://arxiv.org/abs/1308.3581}{{\ttfamily 1308.3581}}].

\bibitem{Klimcik:2014bta}
C.~Klim\v{c}\'{i}k, \emph{{Integrability of the bi-Yang-Baxter sigma-model}},
  \href{https://doi.org/10.1007/s11005-014-0709-y}{\emph{Lett. Math. Phys.}
  {\bfseries 104} (2014) 1095}
  [\href{https://arxiv.org/abs/1402.2105}{{\ttfamily 1402.2105}}].

\bibitem{Delduc:2015xdm}
F.~Delduc, S.~Lacroix, M.~Magro and B.~Vicedo, \emph{{On the Hamiltonian
  integrability of the bi-Yang-Baxter sigma-model}},
  \href{https://doi.org/10.1007/JHEP03(2016)104}{\emph{JHEP} {\bfseries 03}
  (2016) 104} [\href{https://arxiv.org/abs/1512.02462}{{\ttfamily
  1512.02462}}].

\bibitem{Hoare:2014pna}
B.~Hoare, R.~Roiban and A.~A. Tseytlin, \emph{{On deformations of $AdS_n \times
  S^n$ supercosets}},
  \href{https://doi.org/10.1007/JHEP06(2014)002}{\emph{JHEP} {\bfseries 06}
  (2014) 002} [\href{https://arxiv.org/abs/1403.5517}{{\ttfamily 1403.5517}}].

\bibitem{Fateev:1996ea}
V.~A. Fateev, \emph{{The sigma model (dual) representation for a two-parameter
  family of integrable quantum field theories}},
  \href{https://doi.org/10.1016/0550-3213(96)00256-8}{\emph{Nucl. Phys.}
  {\bfseries B473} (1996) 509}.

\bibitem{Novikov:1982ei}
S.~P. Novikov, \emph{{The Hamiltonian formalism and a many valued analog of
  Morse theory}},
  \href{https://doi.org/10.1070/RM1982v037n05ABEH004020}{\emph{Usp. Mat. Nauk}
  {\bfseries 37N5} (1982) 3}.

\bibitem{Witten:1983tw}
E.~Witten, \emph{{Global Aspects of Current Algebra}},
  \href{https://doi.org/10.1016/0550-3213(83)90063-9}{\emph{Nucl. Phys.}
  {\bfseries B223} (1983) 422}.

\bibitem{Witten:1983ar}
E.~Witten, \emph{{Non-abelian Bosonization in Two-Dimensions}},
  \href{https://doi.org/10.1007/BF01215276}{\emph{Commun. Math. Phys.}
  {\bfseries 92} (1984) 455}.

\bibitem{Kawaguchi:2011mz}
I.~Kawaguchi, D.~Orlando and K.~Yoshida, \emph{{Yangian symmetry in deformed
  WZNW models on squashed spheres}},
  \href{https://doi.org/10.1016/j.physletb.2011.06.007}{\emph{Phys. Lett.}
  {\bfseries B701} (2011) 475}
  [\href{https://arxiv.org/abs/1104.0738}{{\ttfamily 1104.0738}}].

\bibitem{Kawaguchi:2013gma}
I.~Kawaguchi and K.~Yoshida, \emph{{A deformation of quantum affine algebra in
  squashed Wess-Zumino-Novikov-Witten models}},
  \href{https://doi.org/10.1063/1.4880341}{\emph{J. Math. Phys.} {\bfseries 55}
  (2014) 062302} [\href{https://arxiv.org/abs/1311.4696}{{\ttfamily
  1311.4696}}].

\bibitem{Delduc:2014uaa}
F.~Delduc, M.~Magro and B.~Vicedo, \emph{{Integrable double deformation of the
  principal chiral model}},
  \href{https://doi.org/10.1016/j.nuclphysb.2014.12.018}{\emph{Nucl. Phys.}
  {\bfseries B891} (2015) 312}
  [\href{https://arxiv.org/abs/1410.8066}{{\ttfamily 1410.8066}}].

\bibitem{Drinfeld:1985rx}
V.~G. Drinfel'd, \emph{{Hopf algebras and the quantum Yang-Baxter equation}},
  {\emph{Sov. Math. Dokl.} {\bfseries 32} (1985) 254}.

\bibitem{Jimbo:1985zk}
M.~Jimbo, \emph{{A q-difference analog of U(g) and the Yang-Baxter equation}},
  \href{https://doi.org/10.1007/BF00704588}{\emph{Lett. Math. Phys.} {\bfseries
  10} (1985) 63}.

\bibitem{Belavin:1984}
A.~A. Belavin and V.~G. Drinfel'd, \emph{{Triangle equations and simple Lie
  algebras}}, {\emph{Sov. Sci. Rev.} {\bfseries C4} (1984) 93}.

\bibitem{Horne:1991gn}
J.~H. Horne and G.~T. Horowitz, \emph{{Exact black string solutions in
  three-dimensions}},
  \href{https://doi.org/10.1016/0550-3213(92)90536-K}{\emph{Nucl. Phys.}
  {\bfseries B368} (1992) 444}
  [\href{https://arxiv.org/abs/hep-th/9108001}{{\ttfamily hep-th/9108001}}].

\bibitem{Giveon:1991jj}
A.~Giveon and M.~Ro\v{c}ek, \emph{{Generalized duality in curved string
  backgrounds}},
  \href{https://doi.org/10.1016/0550-3213(92)90518-G}{\emph{Nucl. Phys.}
  {\bfseries B380} (1992) 128}
  [\href{https://arxiv.org/abs/hep-th/9112070}{{\ttfamily hep-th/9112070}}].

\bibitem{Kawaguchi:2014qwa}
I.~Kawaguchi, T.~Matsumoto and K.~Yoshida, \emph{{Jordanian deformations of the
  $AdS_5 \times S^5$ superstring}},
  \href{https://doi.org/10.1007/JHEP04(2014)153}{\emph{JHEP} {\bfseries 04}
  (2014) 153} [\href{https://arxiv.org/abs/1401.4855}{{\ttfamily 1401.4855}}].

\bibitem{Matsumoto:2014nra}
T.~Matsumoto and K.~Yoshida, \emph{{Lunin-Maldacena backgrounds from the
  classical Yang-Baxter equation - towards the gravity/CYBE correspondence}},
  \href{https://doi.org/10.1007/JHEP06(2014)135}{\emph{JHEP} {\bfseries 06}
  (2014) 135} [\href{https://arxiv.org/abs/1404.1838}{{\ttfamily 1404.1838}}].

\bibitem{Matsumoto:2014gwa}
T.~Matsumoto and K.~Yoshida, \emph{{Integrability of classical strings dual for
  noncommutative gauge theories}},
  \href{https://doi.org/10.1007/JHEP06(2014)163}{\emph{JHEP} {\bfseries 06}
  (2014) 163} [\href{https://arxiv.org/abs/1404.3657}{{\ttfamily 1404.3657}}].

\bibitem{Matsumoto:2015jja}
T.~Matsumoto and K.~Yoshida, \emph{{Yang-Baxter sigma models based on the
  CYBE}}, \href{https://doi.org/10.1016/j.nuclphysb.2015.02.009}{\emph{Nucl.
  Phys.} {\bfseries B893} (2015) 287}
  [\href{https://arxiv.org/abs/1501.03665}{{\ttfamily 1501.03665}}].

\bibitem{Matsumoto:2015uja}
T.~Matsumoto and K.~Yoshida, \emph{{Schr\"{o}dinger geometries arising from
  Yang-Baxter deformations}},
  \href{https://doi.org/10.1007/JHEP04(2015)180}{\emph{JHEP} {\bfseries 04}
  (2015) 180} [\href{https://arxiv.org/abs/1502.00740}{{\ttfamily
  1502.00740}}].

\bibitem{vanTongeren:2015soa}
S.~J. van Tongeren, \emph{{On classical Yang-Baxter based deformations of the
  $AdS_5 \times S^5$ superstring}},
  \href{https://doi.org/10.1007/JHEP06(2015)048}{\emph{JHEP} {\bfseries 06}
  (2015) 048} [\href{https://arxiv.org/abs/1504.05516}{{\ttfamily
  1504.05516}}].

\bibitem{Vicedo:2015pna}
B.~Vicedo, \emph{{Deformed integrable $\sigma$-models, classical R-matrices and
  classical exchange algebra on Drinfel'd doubles}},
  \href{https://doi.org/10.1088/1751-8113/48/35/355203}{\emph{J. Phys.}
  {\bfseries A48} (2015) 355203}
  [\href{https://arxiv.org/abs/1504.06303}{{\ttfamily 1504.06303}}].

\bibitem{Osten:2016dvf}
D.~Osten and S.~J. van Tongeren, \emph{{Abelian Yang-Baxter deformations and
  TsT transformations}},
  \href{https://doi.org/10.1016/j.nuclphysb.2016.12.007}{\emph{Nucl. Phys.}
  {\bfseries B915} (2017) 184}
  [\href{https://arxiv.org/abs/1608.08504}{{\ttfamily 1608.08504}}].

\bibitem{Metsaev:1998it}
R.~R. Metsaev and A.~A. Tseytlin, \emph{{Type IIB superstring action in $AdS_5
  \times S^5$ background}},
  \href{https://doi.org/10.1016/S0550-3213(98)00570-7}{\emph{Nucl. Phys.}
  {\bfseries B533} (1998) 109}
  [\href{https://arxiv.org/abs/hep-th/9805028}{{\ttfamily hep-th/9805028}}].

\bibitem{Berkovits:1999zq}
N.~Berkovits, M.~Bershadsky, T.~Hauer, S.~Zhukov and B.~Zwiebach,
  \emph{{Superstring theory on $AdS_2 \times S^2$ as a coset supermanifold}},
  \href{https://doi.org/10.1016/S0550-3213(99)00683-5}{\emph{Nucl. Phys.}
  {\bfseries B567} (2000) 61}
  [\href{https://arxiv.org/abs/hep-th/9907200}{{\ttfamily hep-th/9907200}}].

\bibitem{Bena:2003wd}
I.~Bena, J.~Polchinski and R.~Roiban, \emph{{Hidden symmetries of the $AdS_5
  \times S^5$ superstring}},
  \href{https://doi.org/10.1103/PhysRevD.69.046002}{\emph{Phys. Rev.}
  {\bfseries D69} (2004) 046002}
  [\href{https://arxiv.org/abs/hep-th/0305116}{{\ttfamily hep-th/0305116}}].

\bibitem{Magro:2008dv}
M.~Magro, \emph{{The Classical Exchange Algebra of $AdS_5 \times S^5$}},
  \href{https://doi.org/10.1088/1126-6708/2009/01/021}{\emph{JHEP} {\bfseries
  01} (2009) 021} [\href{https://arxiv.org/abs/0810.4136}{{\ttfamily
  0810.4136}}].

\bibitem{Vicedo:2010qd}
B.~Vicedo, \emph{{The classical R-matrix of AdS/CFT and its Lie dialgebra
  structure}}, \href{https://doi.org/10.1007/s11005-010-0446-9}{\emph{Lett.
  Math. Phys.} {\bfseries 95} (2011) 249}
  [\href{https://arxiv.org/abs/1003.1192}{{\ttfamily 1003.1192}}].

\bibitem{Green:1983wt}
M.~B. Green and J.~H. Schwarz, \emph{{Covariant Description of Superstrings}},
  \href{https://doi.org/10.1016/0370-2693(84)92021-5}{\emph{Phys. Lett.}
  {\bfseries B136} (1984) 367}.

\bibitem{Grisaru:1985fv}
M.~T. Grisaru, P.~S. Howe, L.~Mezincescu, B.~Nilsson and P.~K. Townsend,
  \emph{{N=2 Superstrings in a Supergravity Background}},
  \href{https://doi.org/10.1016/0370-2693(85)91071-8}{\emph{Phys. Lett.}
  {\bfseries 162B} (1985) 116}.

\bibitem{Tseytlin:1996hs}
A.~A. Tseytlin, \emph{{On dilaton dependence of type II superstring action}},
  \href{https://doi.org/10.1088/0264-9381/13/6/003}{\emph{Class. Quant. Grav.}
  {\bfseries 13} (1996) L81}
  [\href{https://arxiv.org/abs/hep-th/9601109}{{\ttfamily hep-th/9601109}}].

\bibitem{Cvetic:1999zs}
M.~Cveti\v{c}, H.~Lu, C.~N. Pope and K.~S. Stelle, \emph{{T duality in the
  Green-Schwarz formalism, and the massless / massive IIA duality map}},
  \href{https://doi.org/10.1016/S0550-3213(99)00740-3}{\emph{Nucl. Phys.}
  {\bfseries B573} (2000) 149}
  [\href{https://arxiv.org/abs/hep-th/9907202}{{\ttfamily hep-th/9907202}}].

\bibitem{Wulff:2013kga}
L.~Wulff, \emph{{The type II superstring to order $\theta^4$}},
  \href{https://doi.org/10.1007/JHEP07(2013)123}{\emph{JHEP} {\bfseries 07}
  (2013) 123} [\href{https://arxiv.org/abs/1304.6422}{{\ttfamily 1304.6422}}].

\bibitem{Zarembo:2010sg}
K.~Zarembo, \emph{{Strings on Semisymmetric Superspaces}},
  \href{https://doi.org/10.1007/JHEP05(2010)002}{\emph{JHEP} {\bfseries 05}
  (2010) 002} [\href{https://arxiv.org/abs/1003.0465}{{\ttfamily 1003.0465}}].

\bibitem{Pesando:1998wm}
I.~Pesando, \emph{{The GS type IIB superstring action on $AdS_3 \times S^3
  \times T^4$}},
  \href{https://doi.org/10.1088/1126-6708/1999/02/007}{\emph{JHEP} {\bfseries
  02} (1999) 007} [\href{https://arxiv.org/abs/hep-th/9809145}{{\ttfamily
  hep-th/9809145}}].

\bibitem{Rahmfeld:1998zn}
J.~Rahmfeld and A.~Rajaraman, \emph{{The GS string action on $AdS_3 \times S^3$
  with Ramond-Ramond charge}},
  \href{https://doi.org/10.1103/PhysRevD.60.064014}{\emph{Phys. Rev.}
  {\bfseries D60} (1999) 064014}
  [\href{https://arxiv.org/abs/hep-th/9809164}{{\ttfamily hep-th/9809164}}].

\bibitem{Park:1998un}
J.~Park and S.-J. Rey, \emph{{Green-Schwarz superstring on $AdS_3 \times
  S^3$}}, {\emph{JHEP} {\bfseries 01} (1999) 001}
  [\href{https://arxiv.org/abs/hep-th/9812062}{{\ttfamily hep-th/9812062}}].

\bibitem{Chen:2005uj}
B.~Chen, Y.-L. He, P.~Zhang and X.-C. Song, \emph{{Flat currents of the
  Green-Schwarz superstrings in $AdS_5 \times S^1$ and $AdS_3 \times S^3$
  backgrounds}}, \href{https://doi.org/10.1103/PhysRevD.71.086007}{\emph{Phys.
  Rev.} {\bfseries D71} (2005) 086007}
  [\href{https://arxiv.org/abs/hep-th/0503089}{{\ttfamily hep-th/0503089}}].

\bibitem{Babichenko:2009dk}
A.~Babichenko, B.~Stefa\'{n}ski, jr. and K.~Zarembo, \emph{{Integrability and
  the AdS$_3$/CFT$_2$ correspondence}},
  \href{https://doi.org/10.1007/JHEP03(2010)058}{\emph{JHEP} {\bfseries 03}
  (2010) 058} [\href{https://arxiv.org/abs/0912.1723}{{\ttfamily 0912.1723}}].

\bibitem{Cagnazzo:2012se}
A.~Cagnazzo and K.~Zarembo, \emph{{B-field in AdS$_3$/CFT$_2$ Correspondence
  and Integrability}}, \href{https://doi.org/10.1007/JHEP11(2012)133,
  10.1007/JHEP04(2013)003}{\emph{JHEP} {\bfseries 11} (2012) 133}
  [\href{https://arxiv.org/abs/1209.4049}{{\ttfamily 1209.4049}}].

\bibitem{Delduc:2013qra}
F.~Delduc, M.~Magro and B.~Vicedo, \emph{{Integrable deformation of the $AdS_5
  \times S^5$ superstring action}},
  \href{https://doi.org/10.1103/PhysRevLett.112.051601}{\emph{Phys. Rev. Lett.}
  {\bfseries 112} (2014) 051601}
  [\href{https://arxiv.org/abs/1309.5850}{{\ttfamily 1309.5850}}].

\bibitem{Delduc:2014kha}
F.~Delduc, M.~Magro and B.~Vicedo, \emph{{Derivation of the action and
  symmetries of the $q$-deformed $AdS_5 \times S^5$ superstring}},
  \href{https://doi.org/10.1007/JHEP10(2014)132}{\emph{JHEP} {\bfseries 10}
  (2014) 132} [\href{https://arxiv.org/abs/1406.6286}{{\ttfamily 1406.6286}}].

\bibitem{Hoare:2014oua}
B.~Hoare, \emph{{Towards a two-parameter q-deformation of $AdS_3 \times S^3
  \times M^4$ superstrings}},
  \href{https://doi.org/10.1016/j.nuclphysb.2014.12.012}{\emph{Nucl. Phys.}
  {\bfseries B891} (2015) 259}
  [\href{https://arxiv.org/abs/1411.1266}{{\ttfamily 1411.1266}}].

\bibitem{Hoare:2013pma}
B.~Hoare and A.~A. Tseytlin, \emph{{On string theory on $AdS_3 \times S^3
  \times T^4$ with mixed 3-form flux: tree-level S-matrix}},
  \href{https://doi.org/10.1016/j.nuclphysb.2013.05.005}{\emph{Nucl. Phys.}
  {\bfseries B873} (2013) 682}
  [\href{https://arxiv.org/abs/1303.1037}{{\ttfamily 1303.1037}}].

\bibitem{Babichenko:2014yaa}
A.~Babichenko, A.~Dekel and O.~Ohlsson~Sax, \emph{{Finite-gap equations for
  strings on $AdS_3 \times S^3 \times T^4$ with mixed 3-form flux}},
  \href{https://doi.org/10.1007/JHEP11(2014)122}{\emph{JHEP} {\bfseries 11}
  (2014) 122} [\href{https://arxiv.org/abs/1405.6087}{{\ttfamily 1405.6087}}].

\bibitem{Wulff:2014kja}
L.~Wulff, \emph{{Superisometries and integrability of superstrings}},
  \href{https://doi.org/10.1007/JHEP05(2014)115}{\emph{JHEP} {\bfseries 05}
  (2014) 115} [\href{https://arxiv.org/abs/1402.3122}{{\ttfamily 1402.3122}}].

\bibitem{Wulff:2015mwa}
L.~Wulff, \emph{{On integrability of strings on symmetric spaces}},
  \href{https://doi.org/10.1007/JHEP09(2015)115}{\emph{JHEP} {\bfseries 09}
  (2015) 115} [\href{https://arxiv.org/abs/1505.03525}{{\ttfamily
  1505.03525}}].

\bibitem{Arutyunov:2015qva}
G.~Arutyunov, R.~Borsato and S.~Frolov, \emph{{Puzzles of $\eta$-deformed
  $AdS_5 \times S^5$}},
  \href{https://doi.org/10.1007/JHEP12(2015)049}{\emph{JHEP} {\bfseries 12}
  (2015) 049} [\href{https://arxiv.org/abs/1507.04239}{{\ttfamily
  1507.04239}}].

\bibitem{Borsato:2016ose}
R.~Borsato and L.~Wulff, \emph{{Target space supergeometry of $\eta$ and
  $\lambda$-deformed strings}},
  \href{https://doi.org/10.1007/JHEP10(2016)045}{\emph{JHEP} {\bfseries 10}
  (2016) 045} [\href{https://arxiv.org/abs/1608.03570}{{\ttfamily
  1608.03570}}].

\bibitem{Arutyunov:2015mqj}
G.~Arutyunov, S.~Frolov, B.~Hoare, R.~Roiban and A.~A. Tseytlin, \emph{{Scale
  invariance of the $\eta$-deformed $AdS_5 \times S^5$ superstring, T-duality
  and modified type II equations}},
  \href{https://doi.org/10.1016/j.nuclphysb.2015.12.012}{\emph{Nucl. Phys.}
  {\bfseries B903} (2016) 262}
  [\href{https://arxiv.org/abs/1511.05795}{{\ttfamily 1511.05795}}].

\bibitem{Wulff:2016tju}
L.~Wulff and A.~A. Tseytlin, \emph{{Kappa-symmetry of superstring sigma model
  and generalized 10d supergravity equations}},
  \href{https://doi.org/10.1007/JHEP06(2016)174}{\emph{JHEP} {\bfseries 06}
  (2016) 174} [\href{https://arxiv.org/abs/1605.04884}{{\ttfamily
  1605.04884}}].

\bibitem{Hoare:2015wia}
B.~Hoare and A.~A. Tseytlin, \emph{{Type IIB supergravity solution for the
  T-dual of the $\eta$-deformed $AdS_5 \times S^5$ superstring}},
  \href{https://doi.org/10.1007/JHEP10(2015)060}{\emph{JHEP} {\bfseries 10}
  (2015) 060} [\href{https://arxiv.org/abs/1508.01150}{{\ttfamily
  1508.01150}}].

\bibitem{Araujo:2018rbc}
T.~Araujo, E.~\'{O}~Colg\'{a}in and H.~Yavartanoo, \emph{{Embedding the
  modified CYBE in Supergravity}},
  \href{https://doi.org/10.1140/epjc/s10052-018-6335-6}{\emph{Eur. Phys. J.}
  {\bfseries C78} (2018) 854}
  [\href{https://arxiv.org/abs/1806.02602}{{\ttfamily 1806.02602}}].

\bibitem{Hoare:2018ebg}
B.~Hoare and F.~K. Seibold, \emph{{Poisson-Lie duals of the $\eta$-deformed
  $AdS_2 \times S^2 \times T^6$ superstring}},
  \href{https://doi.org/10.1007/JHEP08(2018)107}{\emph{JHEP} {\bfseries 08}
  (2018) 107} [\href{https://arxiv.org/abs/1807.04608}{{\ttfamily
  1807.04608}}].

\bibitem{Klimcik:1995ux}
C.~Klim\v{c}\'{i}k and P.~\v{S}evera, \emph{{Dual Non-Abelian duality and the
  Drinfel'd double}},
  \href{https://doi.org/10.1016/0370-2693(95)00451-P}{\emph{Phys. Lett.}
  {\bfseries B351} (1995) 455}
  [\href{https://arxiv.org/abs/hep-th/9502122}{{\ttfamily hep-th/9502122}}].

\bibitem{Klimcik:1995jn}
C.~Klim\v{c}\'{i}k, \emph{{Poisson-Lie T duality}},
  \href{https://doi.org/10.1016/0920-5632(96)00013-8}{\emph{Nucl. Phys. Proc.
  Suppl.} {\bfseries 46} (1996) 116}
  [\href{https://arxiv.org/abs/hep-th/9509095}{{\ttfamily hep-th/9509095}}].

\bibitem{Klimcik:1995dy}
C.~Klim\v{c}\'{i}k and P.~\v{S}evera, \emph{{Poisson-Lie T duality and loop
  groups of Drinfel'd doubles}},
  \href{https://doi.org/10.1016/0370-2693(96)00025-1}{\emph{Phys. Lett.}
  {\bfseries B372} (1996) 65}
  [\href{https://arxiv.org/abs/hep-th/9512040}{{\ttfamily hep-th/9512040}}].

\bibitem{Klimcik:2017ken}
C.~Klim\v{c}\'{i}k, \emph{{Yang-Baxter $\sigma$-model with WZNW term as
  ${\mathcal E}$-model}},
  \href{https://doi.org/10.1016/j.physletb.2017.07.051}{\emph{Phys. Lett.}
  {\bfseries B772} (2017) 725}
  [\href{https://arxiv.org/abs/1706.08912}{{\ttfamily 1706.08912}}].

\bibitem{Demulder:2017zhz}
S.~Demulder, S.~Driezen, A.~Sevrin and D.~C. Thompson, \emph{{Classical and
  Quantum Aspects of Yang-Baxter Wess-Zumino Models}},
  \href{https://doi.org/10.1007/JHEP03(2018)041}{\emph{JHEP} {\bfseries 03}
  (2018) 041} [\href{https://arxiv.org/abs/1711.00084}{{\ttfamily
  1711.00084}}].

\bibitem{Severa:2018pag}
P.~\v{S}evera and F.~Valach, \emph{{Courant algebroids, Poisson-Lie T-duality,
  and type II supergravities}},
  \href{https://arxiv.org/abs/1810.07763}{{\ttfamily 1810.07763}}.

\bibitem{Demulder:2018lmj}
S.~Demulder, F.~Hassler and D.~C. Thompson, \emph{{Doubled aspects of
  generalised dualities and integrable deformations}},
  \href{https://arxiv.org/abs/1810.11446}{{\ttfamily 1810.11446}}.

\bibitem{Delduc:2016ihq}
F.~Delduc, S.~Lacroix, M.~Magro and B.~Vicedo, \emph{{On q-deformed symmetries
  as Poisson-Lie symmetries and application to Yang-Baxter type models}},
  \href{https://doi.org/10.1088/1751-8113/49/41/415402}{\emph{J. Phys.}
  {\bfseries A49} (2016) 415402}
  [\href{https://arxiv.org/abs/1606.01712}{{\ttfamily 1606.01712}}].

\bibitem{Arutyunov:2013ega}
G.~Arutyunov, R.~Borsato and S.~Frolov, \emph{{S-matrix for strings on
  $\eta$-deformed $AdS_5 \times S^5$}},
  \href{https://doi.org/10.1007/JHEP04(2014)002}{\emph{JHEP} {\bfseries 04}
  (2014) 002} [\href{https://arxiv.org/abs/1312.3542}{{\ttfamily 1312.3542}}].

\bibitem{Regelskis:2015xxa}
V.~Regelskis, \emph{{Yangian of AdS$_3$/CFT$_2$ and its deformation}},
  \href{https://doi.org/10.1016/j.geomphys.2016.04.001}{\emph{J. Geom. Phys.}
  {\bfseries 106} (2016) 213}
  [\href{https://arxiv.org/abs/1503.03799}{{\ttfamily 1503.03799}}].

\bibitem{Beisert:2008tw}
N.~Beisert and P.~Koroteev, \emph{{Quantum Deformations of the One-Dimensional
  Hubbard Model}},
  \href{https://doi.org/10.1088/1751-8113/41/25/255204}{\emph{J. Phys.}
  {\bfseries A41} (2008) 255204}
  [\href{https://arxiv.org/abs/0802.0777}{{\ttfamily 0802.0777}}].

\bibitem{Borsato:2012ud}
R.~Borsato, O.~Ohlsson~Sax and A.~Sfondrini, \emph{{A dynamic
  $\mathfrak{su}(1|1)^2$ S-matrix for AdS$_3$/CFT$_2$}},
  \href{https://doi.org/10.1007/JHEP04(2013)113}{\emph{JHEP} {\bfseries 04}
  (2013) 113} [\href{https://arxiv.org/abs/1211.5119}{{\ttfamily 1211.5119}}].

\bibitem{Borsato:2013qpa}
R.~Borsato, O.~Ohlsson~Sax, A.~Sfondrini, B.~Stefa\'{n}ski, jr. and
  A.~Torrielli, \emph{{The all-loop integrable spin-chain for strings on $AdS_3
  \times S^3 \times T^4$: the massive sector}},
  \href{https://doi.org/10.1007/JHEP08(2013)043}{\emph{JHEP} {\bfseries 08}
  (2013) 043} [\href{https://arxiv.org/abs/1303.5995}{{\ttfamily 1303.5995}}].

\bibitem{Borsato:2014hja}
R.~Borsato, O.~Ohlsson~Sax, A.~Sfondrini and B.~Stefa\'{n}ski, jr., \emph{{The
  complete $AdS_3 \times S^3 \times T^4$ worldsheet S matrix}},
  \href{https://doi.org/10.1007/JHEP10(2014)066}{\emph{JHEP} {\bfseries 10}
  (2014) 66} [\href{https://arxiv.org/abs/1406.0453}{{\ttfamily 1406.0453}}].

\bibitem{Lacroix:2018njs}
S.~Lacroix, \emph{{Integrable models with twist function and affine Gaudin
  models}}, Ph.D. thesis, ENS de Lyon, 2018
\newblock [\href{https://arxiv.org/abs/1809.06811}{{\ttfamily 1809.06811}}].

\bibitem{Vicedo:2017cge}
B.~Vicedo, \emph{{On integrable field theories as dihedral affine Gaudin
  models}}, \href{https://doi.org/10.1093/imrn/rny128}{\emph{Int. Math. Res.
  Not. IMRN} (2018) rny128} [\href{https://arxiv.org/abs/1701.04856}{{\ttfamily
  1701.04856}}].

\end{thebibliography}
\end{document}